\def\la{\langle}
\def\ra{\rangle}
\def\lb{\left\{}
\def\rb{\right\}}
\def\rf#1{(\ref{#1})}
\def\xsb#1{Subsection~\ref{#1}}
\def\xsc#1{Section~\ref{#1}}
\def\xap#1{Appendix~\ref{#1}}
\def\xfg#1{Figure~\ref{#1}}
\def\bgm{\mbox{\boldmath$\gamma$}}
\def\bzt{\mbox{\boldmath$\zeta$}}
\def\et{{ \it et al.}}
\def\tld#1{\tilde{\bf#1}}
\def\xht#1{\widehat{\bf#1}}
\def\cu{{\it Ch1}{\kern .035em}}
\def\cd{{\it Ch1a}{\kern .035em}}
\def\ct{{\it Ch2}{\kern .035em}}
\documentstyle[12pt,psfig]{article}
\newif
\iffigs
\figstrue   
\makeatletter \def\fps@figure{tp} \makeatother
\iffigs
\fi
\def\drawing #1 #2 #3 {
\begin{center}
\setlength{\unitlength}{1mm}
\begin{picture}(#1,#2)(0,0)
\put(0,0){\framebox(#1,#2){#3}}
\end{picture}
\end{center} }

\begin{document}

{\center
{\it Short title: LARGE-SCALE MAGNETIC FIELD INSTABILITY}

\bigskip
{\bf Dynamo effect in parity-invariant flow with large and moderate
separation of scales}

\bigskip
V.A. Zheligovsky$^{a,b,c,}$\footnote{E-mail: vlad@mitp.ru},
O.M. Podvigina$^{a,b,c,}$\footnote{E-mail: olgap@mitp.ru},
U. Frisch$^{a,}$\footnote{E-mail: uriel@obs-nice.fr}

\bigskip
$^a$Observatoire de la C\^ote d'Azur, CNRS UMR~6529,\\
BP~4229, 06304 Nice Cedex 4, France

\bigskip
$^b$International Institute of Earthquake Prediction Theory\\
and Mathematical Geophysics,\\
79 bldg.~2, Warshavskoe ave., 113556 Moscow, Russian Federation

\bigskip
$^c$Laboratory of general aerodynamics, Institute of Mechanics,\\
Lomonosov Moscow State University,\\
1, Michurinsky ave., 119899 Moscow, Russian Federation

\bigskip
{\it Submitted to Geophysical Astrophysical Fluid Dynamics}\\
 \centerline{Revised version, April 2001}
}

\begin{abstract}\noindent
It is shown that non-helical (more precisely, parity-invariant) flows
capable of sustaining a large-scale dynamo by the negative
magnetic eddy diffusivity effect are quite common. This conclusion is
based on numerical examination of a large number of randomly selected
flows. Few outliers with strongly negative eddy diffusivities are
also found, and they are interpreted in terms of the closeness of the
control parameter to a critical value for generation of a small-scale
magnetic field. Furthermore, it is shown that, for
parity-invariant flows, a moderate separation of scales between the
basic flow and the magnetic field often significantly reduces the
critical magnetic Reynolds number for the onset of dynamo action.

\end{abstract}

\newpage
\section{Introduction}
\label{s:intro}

The kinematic dynamo effect is generally believed to be facilitated by
a separation of spatial scales between the flow and the
magnetic field. This idea has its roots in the discovery of the
$\alpha$-effect (Steenbeck\et, 1966; Krause and R\"adler, 1980) for
non-parity-invariant flows (an instance of which are helical
flows). Indeed, recent results indicate that helicity and
scale separation can lead to experimental dynamos
(Gailitis\et, 2000; Busse, 2000, Stieglitz and M\"uller, 2001).

Lanotte\et\ (1999) showed that scale separation can also be useful for
flows possessing parity invariance and thus lacking
$\alpha$-effect. They found a first instance of a three-dimensional
parity-invariant flow, the ``modified Taylor--Green flow'', which can
amplify a large-scale magnetic field by a negative eddy diffusivity
mechanism, as conjectured by Kraichnan (1976). Examples of negative
eddy diffusivities when the magnetic field depends on all three
coordinates, while the flow depends only on two (two and a half
dimensional dynamos) were actually known since the work of Roberts
(1972).  New instances of large-scale dynamos with the Roberts flow
were found more recently by Tilgner (1997) and 
Plunian and R\"adler (2000, 2001). Emergence of
the $\alpha$-effect and the negative eddy diffusivity effect can be
demonstrated by asymptotic theory involving, in principle, very large
scale separation. Both the aforementioned experiments and numerical
simulations with helical (Galanti\et, 1992) flows indicate that even a
moderate scale separation of a factor two can help for obtaining a
dynamo (a physical interpretation for the case of the ABC flows has
been proposed by Archontis (2000)).

The following argument due to Fauve (1999) shows, how crucial for
experimental dynamos it is to lower the critical magnetic Reynolds
number $R_m=LV/\eta$. Here, $L$ is the overall scale of the apparatus,
$V$ is the typical velocity and $\eta$ is the (molecular) magnetic
diffusivity. Given the typically very small value of the magnetic
Prandtl number, the flow is expected to be strongly turbulent. Hence,
the kinetic energy dissipated per unit time in the apparatus is
$P\sim\rho(V^3/\ell)L^3$, where $\rho$ is the fluid density and $\ell$
is the typical spatial scale of the velocity. Thus, $P\sim\rho\eta ^3
R_m^3/(\epsilon L)$, where $\epsilon=\ell/L$ is the scale-separation
factor. Therefore, for a given fluid ($\eta$), geometry ($\epsilon$) and
size ($L$), the required pump power behaves as the cube of the
attainable $R_m$. It is generally believed that the cost of an
experiment scales roughly as its volume, i.e. $\sim L^3$; thus, if
there are practical limitations on the maximum pump power available,
the cost will scale as the ninth power of the attainable $R_m$. All
this makes it desirable to design apparatus with the lowest possible
critical magnetic Reynolds number.

Our intention is to begin a systematic investigation of the kinematic
dynamo effect for parity-invariant flows with large or moderate scale
separation with emphasis on (i) the genericity of the phenomenon of
negative eddy diffusivity and (ii) the lowering of the critical
magnetic Reynolds number. Note, that parity-invariant flows are at
least as ``natural'' as helical ones. Indeed, the Navier-Stokes
equations are parity-invariant. Hence, even if this particular
symmetry, like most other symmetries, is broken by the mechanisms
generating the flow, we can expect the symmetry to be dynamically
restored in a statistical sense at high Reynolds numbers (Frisch,
1995). Note also that the existence of three-dimesnional flow 
with negative eddy diffusivity is  a recent discovery; 
contrary to the alpha effect which can be derived in purely analytic 
fashion  by working, e.g., at
low magnetic Reynolds numbers or with a delta-correlated (in time)
velocity, only (very CPU-demanding) numerical solutions of the 
equations arising from  multi-scale techniques can demonstrate
the existence of negative eddy-diffusivities and can show that
it is not a pathologically unusual effect. 
We shall work here with time-independent flows, which we may
think of as being averages of turbulent flows. We are of course aware
that time dependence of the flow and, generally more realistic choices 
of the basic flow could somewhat modify our conclusions;  such 
issues are left for future work.

The paper is organized as follows.

In Sections~\ref{s:expansion} and \ref{s:abundance} we
consider the case where the spatial scale of magnetic field is much
larger than that of the flow. Then $\epsilon$ is a natural small
parameter, enabling one to apply asymptotic methods. Multiscale
expansions in hydrodynamics may be viewed as adaptations of general
homogenization methods used to determine effective large-scale
transport coefficients of periodic or random materials (Bensoussan\et,
1978; Kozlov, 1978; Zhikov\et, 1979). Usually the effective
coefficients cannot be obtained without solving suitable auxiliary
small-scale problems. In three space dimensions this may require
significant computational resources. In Lanotte\et\ (1999)
equations for the large-scale component of magnetic fields are derived
along the same lines as in previous works on negative eddy
viscosity in ordinary incompressible hydrodynamics (Dubrulle and
Frisch, 1991; Gama\et, 1994; Wirth\et, 1995). Here we use a similar but
a distinct approach based on the work of Vishik (1986, 1987), which yields
a complete decomposition of magnetic modes and their eigenvalues (in
\xap{a:expansion} arbitrarily high-order terms of such expansions are derived).

To investigate, how commonly the phenomenon of negative magnetic eddy
diffusivity is exhibited by fully  three-dimensional flows, we follow in
\xsc{s:abundance} a strategy similar to the one, which Gama\et\ (1994)
applied to estimate the abundance of two-dimensional flows with negative
eddy viscosity. By solving numerically the relevant auxiliary
problems we determine the minimum over directions of the (anisotropic)
magnetic eddy diffusivities for two ensembles of flows with random
Fourier components. This way we determine the fraction of flows such
that, when decreasing molecular magnetic diffusivity, a
large-scale negative eddy diffusivity instability appears before the
flow becomes a small-scale dynamo. An interpretation of ``outliers'' with
strongly negative eddy diffusivity, found in computations, is given in
\xap{a:loweddydiff}.
Since it is important for our purpose to analyze as many instances as
possible, we have used a new efficient algorithm for solving the
elliptic auxiliary problems arising from the two-scale expansion. This
method is an outgrowth of the Chebyshev iterative method developed by
Podvigina and Zheligovsky (1997) for self-adjoint negative definite
linear operators. The operators arising in hydrodynamical problems are
not self-adjoint, but they naturally split into a dominant
self-adjoint negative definite part (the molecular diffusion term) and
the remaining transport term. We can then combine Chebyshev
methods with operator splitting to reduce the harm stemming from
complex eigenvalues. This method and its theoretical foundations are
presented in detail in Zheligovsky (2001), available only in
Russian. For the reader's convenience, the essentials are given in
\xap{a:chebyshev}.

In \xsc{s:moderatescale} we turn to cases with moderate scale
separation (i.e. when $\epsilon$ is finite). First, we consider
instances of flows for which we already know that there is a negative
magnetic eddy diffusivity, and we determine windows in $\epsilon$ over
which dynamo action is present. Second, we examine the case
$\epsilon=1/2$, that is, we assume that the magnetic field has a
spatial period twice that of the flow, and study how the critical
magnetic Reynolds number $R_m$ is affected, compared to the case with
no scale separation ($\epsilon=1$). In both types of computations we
exploit the fact that, for space-periodic boundary conditions,
eigenmodes have the form
\begin{equation}
{\bf H}=e^{i\epsilon\bf q\cdot x}{\bf h}({\bf x};\epsilon,{\bf q}),
\label{Hform}\end{equation}
where $\bf q$ is a prescribed wavevector. This enables one to solve a
problem involving only small scales, but parameterized by $\epsilon$
and $\bf q$ (cf.~Roberts, 1972; Frisch\et, 1986). Solution of the
eigenvalue problem is more CPU-intensive than the solution of
auxiliary problems involved in the asymptotic case (because efficient
preconditioning techniques are available for the latter). Numerical
methods and codes used for the eigenvalue computations are adapted
from Zheligovsky (1993).

In \xsc{s:conclusion} we summarize our main results.

\section{Asymptotic expansion of magnetic modes and of their eigenvalues}
\label{s:expansion}

Here we briefly review the asymptotic expansion arising in the
kinematic dynamo problem, when the scale ratio $\epsilon$ is small.
The conducting fluid flow $\bf v(x)$ is assumed to be
time-independent, and the problem therefore reduces to the eigenvalue
problem for the magnetic induction operator
\begin{equation}
\eta\nabla^2{\bf H}+\nabla\times({\bf v}\times{\bf H})=\lambda{\bf H}
\label{eigeneq}\end{equation}
in the space of solenoidal vector fields:
\begin{equation}
\nabla\cdot{\bf H}=0.
\label{Hsolenoidality}
\end{equation}
Here $\eta>0$ is magnetic molecular diffusivity, ${\bf H}({\bf x})$
is a magnetic mode and $\lambda$ is the associated eigenvalue
(the real part of which is here referred to as the growth rate of the mode;
negative growth rates correspond to decaying modes).
The flow is assumed incompressible:
\begin{equation}
\nabla\cdot{\bf v}=0,
\label{vsolenoidality}\end{equation}
$2\pi$-periodic in space and parity-invariant:
\begin{equation}
{\bf v}({\bf x})=-{\bf v}(-{\bf x}).
\label{vparityinv}\end{equation}

The magnetic mode ${\bf H}({\bf x},{\bf y})$ is supposed to depend
on the fast variable $\bf x$ and on the
slow variable $\bf y=\epsilon x$, and is $2\pi$-periodic in both fast
and slow variables (the velocity $\bf v$ is independent of slow variables).
Introduction of the two spatial scales results in a ``decomposition
of the gradient'' $\nabla\to\nabla_{\bf x}+\epsilon\nabla_{\bf y}$
in \rf{eigeneq} and \rf{Hsolenoidality}.
Here and in what follows the subscripts $\bf x$ and $\bf y$ refer
to differential operators in fast and slow variables, respectively.

Let $\la\cdot\ra$ and $\lb\cdot\rb$ denote the mean (over the
periodicity cube) and the fluctuating part of a vector field, respectively:
\begin{equation}
\la{\bf f}\ra=(2\pi)^{-3}\int_{T^3}{\bf f}({\bf x},{\bf y})d{\bf x},\quad
\lb{\bf f}\rb={\bf f}-\la{\bf f}\ra.
\label{defmeanosc}\end{equation}
Parity invariance \rf{vparityinv} implies $\la{\bf v}\ra=0$.

Solutions to the eigenproblem \rf{eigeneq}-\rf{vparityinv} are sought
in the form of asymptotic series
\begin{equation}
{\bf H}=\sum_{n=0}^\infty{\bf h}_n({\bf x},{\bf y})\epsilon^n,
\label{Hseries}\end{equation}
\begin{equation}
\lambda=\sum_{n=0}^\infty\lambda_n\epsilon^n.
\label{lambdaseries}\end{equation}
As shown in \xap{a:expansion}, substitution of these expansions in \rf{eigeneq}
leads to a hierarchy of equations. By application of standard techniques
for singular perturbation expansions, terms in the expansion eigenvalues are
successively determined from solvability conditions, obtained here by just
averaging the equations. In particular, it can be shown that
$\lambda_0=\lambda_1=0$, and the leading term $\lambda_2$ in the series
\rf{lambdaseries} can be determined from the following eigenvalue problem
\begin{equation}
\eta\nabla^2_{\bf y}\la{\bf h}_0\ra+\nabla_{\bf y}\times
\sum_{k=1}^3\sum_{m=1}^3{\bf D}_{mk}{\partial\la h^k_0\ra
\over\partial y_m}=\lambda_2\la{\bf h}_0\ra,
\label{leadingtermseq}\end{equation}
subject to the solenoidality condition
\begin{equation}
\nabla_{\bf y}\cdot\la{\bf h}_0\ra=0.
\label{h0solenoidality}\end{equation}

Thus, the leading terms of the expansions in powers of $\epsilon$ of
the mean magnetic field and of the eigenvalue turn out to be,
respectively, an eigenfunction and the associated eigenvalue of a
partial differential operator comprised only of second order derivatives,
which may be regarded as representing an anisotropic
diffusion. However, this operator is not necessarily dissipative: it
may have eigenvalues with positive real parts. In this context one
then may speak of negative eddy diffusivity.

As shown in \xap{a:expansion} eqn.~\rf{wasnotnumbered}, the coefficients 
${\bf D}_{mk}$ of the homogenised  equation \rf{leadingtermseq} are given by
\begin{equation}
{\bf D}_{mk}=\la{\bf v}\times{\bf\Gamma}_{mk}\ra,
\label{difftensor}\end{equation}
where the tensor field ${\bf\Gamma}_{mk}$ is obtained by solving
a set of two auxiliary problems:
\begin{equation}
\eta\nabla^2{\bf S}_k+\nabla\times({\bf v}\times{\bf S}_k)=
-{\partial{\bf v}\over\partial x_k}
\label{Seq}\end{equation}
and
\begin{equation}
\eta\nabla^2{\bf\Gamma}_{mk}+\nabla\times({\bf v}\times{\bf\Gamma}_{mk})=
-2\eta{\partial{\bf S}_k\over\partial x_m}
+v^m({\bf S}_k+{\bf e}_k)-{\bf v}S^m_k,
\label{Gammaeq}\end{equation}
which will be referred to as the first and the second auxiliary problem,
respectively. Here, ${\bf e}_k$ is the unit vector in the $k$-th
cartesian coordinate direction and superscripts denote cartesian
components of vectors. For solutions to \rf{Seq} and \rf{Gammaeq} to
exist, we require that zero should not be an eigenvalue of the
magnetic induction operator
$$
{\cal L}{\bf H}\equiv\eta\nabla^2{\bf H}+\nabla\times({\bf v}\times{\bf H}),
$$
restricted to small-scale solenoidal vector fields with
a vanishing mean; by ``small-scale'' we mean independent of the
slow variable. Generically, this condition is satisfied.

Problems \rf{Seq} and \rf{Gammaeq} are formulated with a
$2\pi$-periodicity boundary condition for the vector fields
${\bf S}_k({\bf x})$ and ${\bf\Gamma}_{mk}({\bf x})$, which should also have
a vanishing mean. It follows that the ${\bf S}_k$'s are
parity-antiinvariant \hbox{(${\bf S}_k({\bf x})={\bf S}_k(-{\bf x})$~)} and
solenoidal ($\nabla\cdot{\bf S}_k=0$), and that the
${\bf\Gamma}_{mk}$'s are parity-invariant
(${\bf\Gamma}_{mk}({\bf x})=-{\bf\Gamma}_{mk}(-{\bf x})$~).

Since the homogenized operator defined by
\rf{leadingtermseq}-\rf{h0solenoidality} has constant
coefficients, its eigenvectors are Fourier harmonics:
$\la{\bf h}_0\ra={\bf h}e^{i\bf q\cdot y}$, where $\bf h$ and $\bf q$ are
constant vectors. The wavevector $\bf q$ is taken real because we are
only interested in modes uniformly bounded in the entire space. Substitution
in \rf{leadingtermseq}-\rf{h0solenoidality} yields
\begin{equation}
\eta|{\bf q}|^2{\bf h}+{\bf q}\times\sum_{k=1}^3\sum_{m=1}^3
{\bf D}_{mk}h^kq^m=-\lambda_2{\bf h},
\label{3deigeneq}\end{equation}
\begin{equation}
{\bf h}\cdot{\bf q}=0.
\label{3dsolenoidality}
\end{equation}

\section{Abundance of flows producing negative magnetic eddy diffusivity}
\label{s:abundance}

Following Gama\et\ (1994), we now ask how common is the phenomenon of
negative diffusivity.

Time-independent periodic flows, satisfying \rf{vsolenoidality}-\rf{vparityinv},
are represented here by their Fourier series
\begin{equation}
{\bf v}=\sum_{|{\bf k|}=1}^N{\bf v}_{\bf k}e^{i{\bf kx}}
\label{vFourier}
\end{equation}
with random amplitude harmonics. For the velocity to be real,
${\bf v_k}=\overline{\bf v}_{-{\bf k}}$, and parity invariance \rf{vparityinv}
requires Re\,$\bf v_k$=0.

We study two ensembles of flows with different types of fall-offs of
the energy spectrum $E_K$. The latter is defined as the total energy
contained in harmonics, whose wavevectors $\bf k$ belong to the $K$-th
spherical shell, i.e. $K-1<|{\bf k}|\le K$, where $K>0$ is integer.
When $E_K\propto e^{-\xi K}$, the flow is said to have exponential
spectrum; when $E_K\propto K^{-1}$, it is said to have hyperbolic
spectrum. These have respectively rapid and slow fall-offs. The
spectra are band-limited and set to zero beyond wavenumber $N=10$ for
the exponential case and beyond $N=7$ for the hyperbolic case. The value
$\xi=10^{-2/3}$ is used throughout in the exponential case.

The flows are produced by the following procedure (applied only to one
half of the set of wavevectors, the second one being related by complex
conjugation). For $\bf k$ in the ball $1\le|{\bf k}|\le N$, random
imaginary vectors $\bf r_k$ are generated with the imaginary part of
each cartesian component uniformly distributed in the interval
$[-0.5,0.5]$ (the RAN2 random number generator by Press\et, 1992, is
used). Then the vectors $\bf r_k$ are projected onto the plane
perpendicular to $\bf k$ to gain incompressibility. Finally, the
obtained vectors are normalized, so that the flow has the desired
energy spectrum and the root-mean-square velocity is one, when the normalized
vectors are employed as the Fourier coefficients $\bf v_k$ in \rf{vFourier}.

Each ensemble consists of 100 flows. For each flow, computations are
carried out for three values of molecular diffusivity, $\eta=0.3$~,
0.2 and 0.1~. Note that when $\eta$ is too large, no dynamo action can
take place at large or small scales because the eddy diffusivity will
basically be equal to the molecular diffusitivity. (In contrast, an
alpha-effect dynamo is still possible if the system is large
enough). If $\eta$ is too small, small-scale dynamo instabilities will
have much larger growth rates than large-scale instabilities. In both
extreme cases it is meaningless to study negative eddy diffusivity
large-scale instabilities. The range of values of $\eta$
chosen, between 0.1 and 0.3, is an empirically
determined tradeoff between these extremes. All computations
are performed with the resolution of $64^3$ Fourier harmonics,
employing standard spectral methods based on fast Fourier transforms
and dealiasing. (Each case requires from 10 to 30~minutes of CPU on
one DEC Alpha~EV6 processor, depending on the value of $\eta$.) The
power spectra of the solutions to \rf{Seq} and
\rf{Gammaeq} turn out to decrease monotonically very quickly for
$K\ge2$. For $\eta=0.1$ in the hyperbolic case (resp. exponential
case), the decrease is typically by 7-8 orders of magnitude (14-15
orders of magnitude) from maximum to minimum. Note that our code
reproduces the negative eddy diffusivity values obtained by an
alternative numerical method by Lanotte\et\ (1999) for the modified
Taylor--Green flow.

\begin{figure}
\iffigs
\centerline{\psfig{file=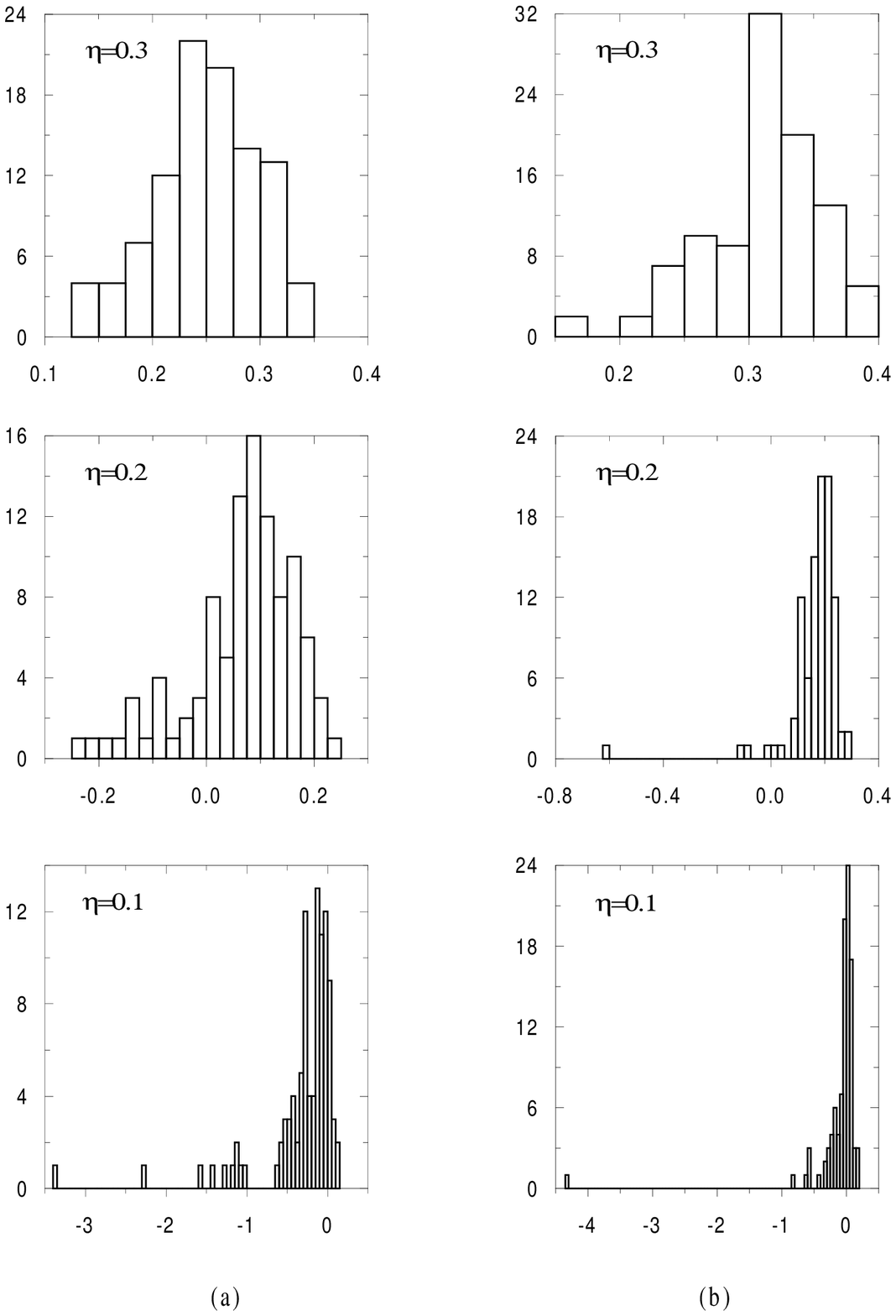,width=116mm,clip=}}
\else
\drawing 100 10 {eddy diffusivity histograms}
\fi
\caption{Histograms of the minimum eddy diffusivity values $\eta_{\rm eddy}$ (horizontal axis)
for ensembles of 100 parity-invariant flows with exponential (a) and
hyperbolic (b) spectra.}
\label{f:eddyhisto}
\end{figure}

\begin{table}
Table 1. Statistics of the eddy diffusivities
for the considered ensembles of flows.

\bigskip
\begin{tabular}{|c|c|c|c|c|c|c|}\hline
&\multicolumn{3}{|c|}{Exponential spectrum flows}&
\multicolumn{3}{|c|}{Hyperbolic spectrum flows}\\ \cline{2-7}
&$\eta=0.3$&$\eta=0.2$&$\eta=0.1$&$\eta=0.3$&$\eta=0.2$&$\eta=0.1$\\ \hline
Re\,$\eta_{\rm eddy}<0$&0\%&18\%&86\%&0\%&4\%&53\%\\
Re\,$\eta_{\rm eddy}<\eta$&83\%&96\%&98\%&30\%&63\%&94\%\\
complex $\eta_{\rm eddy}$&7\%&2\%&2\%&8\%&5\%&3\%\\ \hline
\end{tabular}
\end{table}

After the homogenized coefficients \rf{difftensor} are determined, the
minimum magnetic eddy diffusivity
$$
\eta_{\rm eddy}=\min_{|{\bf q}|=1}(-\lambda_2({\bf q})\ )
$$
is calculated as the minimum eigenvalue of the problem
\rf{3deigeneq}-\rf{3dsolenoidality}.
(Note, that for a unit wavevector the eigenvalue is just
the opposite of the eddy diffusivity.) For a given $\bf q$ this
eigenproblem yields two eigenvalues satisfying a quadratic equation;
the eigenvalues can thus be complex conjugate (a similar phenomenon was
observed for three-dimensional eddy viscosities; see Wirth\et,
1995). Then the minimum of the real part of the complex eddy
diffusivity, which determines the maximum rate of instability of a
large-scale magnetic perturbation, is sought. However, in our computations
complex eddy diffusivities appear quite rarely.

Histograms of the obtained magnetic eddy diffusivities are shown in
\xfg{f:eddyhisto}. The statistics of eddy diffusivities shown
in Table~1 indicate that the lower the molecular diffusivity, the higher the 
chance that a flow will have  a negative ``eddy correction''  (correction to 
the molecular value coming from the flow) and also the higher the chance
that the resulting eddy viscosity will be negative. Note however that 
the eddy correction (given, at least implicitly, by 
\rf{difftensor}-\rf{Gammaeq}) has a complicated, generally
non-monotonic, nonlinear dependence on the molecular diffusivity. 
Table~1 also shows that
flows with the exponential energy spectrum
are more efficient in lowering the minimum of the eddy diffusivity than those
with the hyperbolic spectrum. This suggests that low
wavenumber harmonics of the underlying flow are more important for
lowering than high wavenumber harmonics.

\begin{figure}[p]
\iffigs
\centerline{\psfig{file=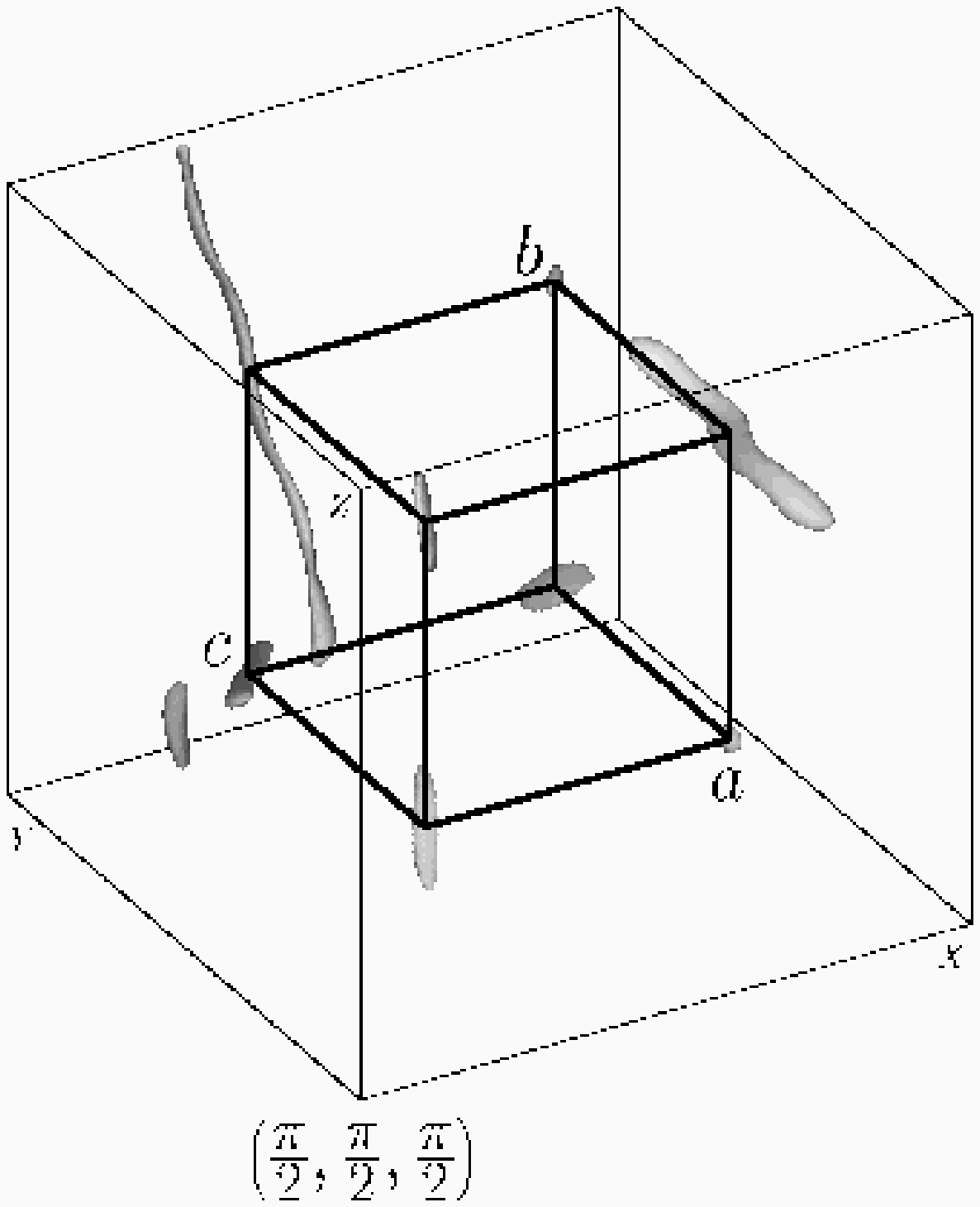,width=7cm,clip=}}
\vspace{-1truemm}
\centerline{\psfig{file=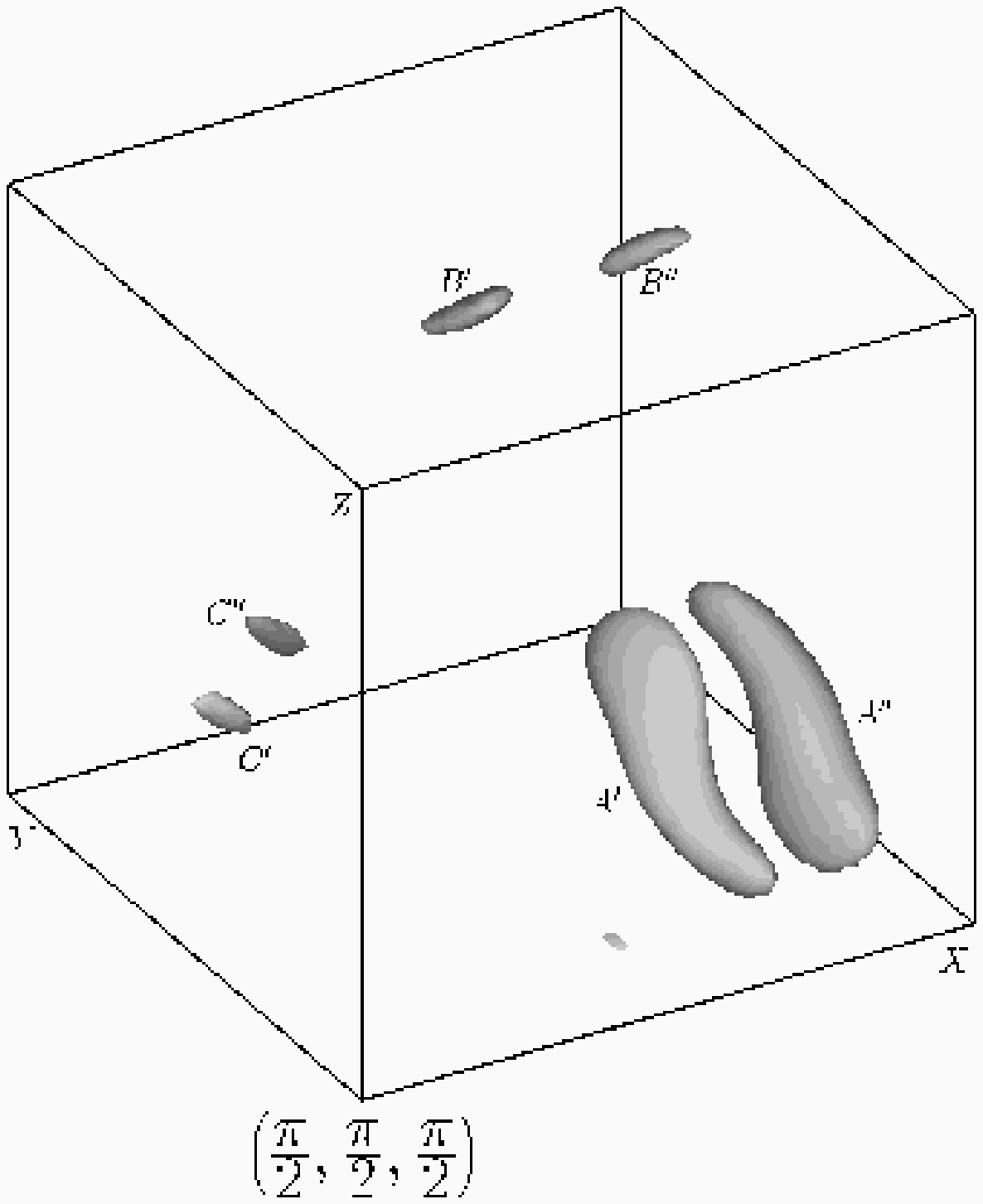,width=7cm,clip=}
\psfig{file=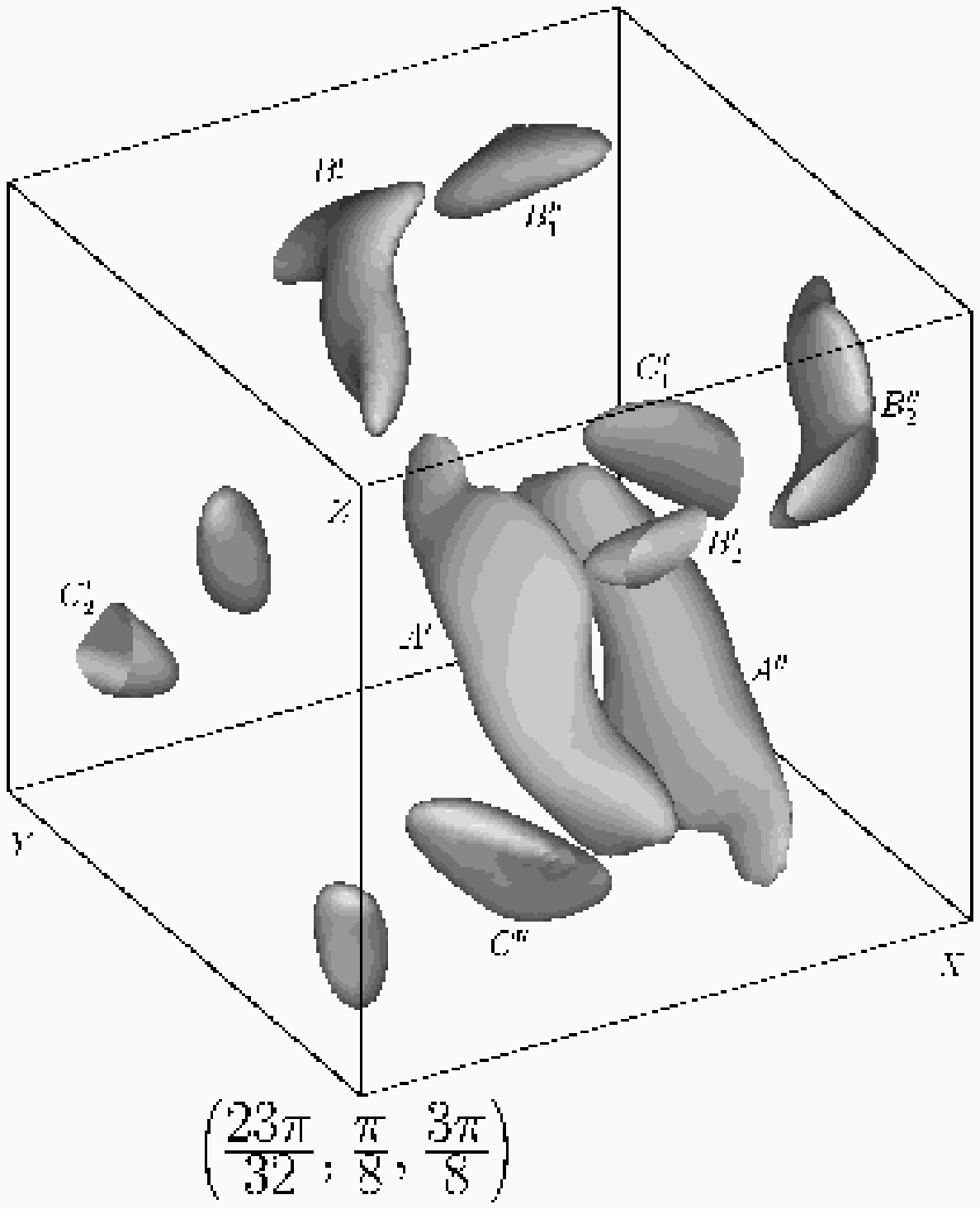,width=7cm,clip=}}
\else
\drawing 100 10 {eigenmode sample 1 (exponential spectrum)}
\fi
\caption{First sample of velocity field and
corresponding magnetic eigenfunction shown as isosurfaces. Upper:
$|{\bf v}|^2$ at 0.2\% of its maximum; lower left: $|{\bf H}|^2$ at
60\% of its maximum; lower right: $|{\bf H}|^2$ at 40\% of its
maximum. Vertices of the bold solid line cube are stagnation points; the
relevant ones are labelled $a$, $b$ and $c$, the associated magnetic
stuctures being labelled with the corresponding capital letters;
primes and double-primes distinguish pairs of structures; subscripts
such as 1 or 2 indicate the continuation of a given structure by
periodicty.}
\label{f:sample1}
\end{figure}

\begin{figure}[p]
\iffigs
\centerline{\psfig{file=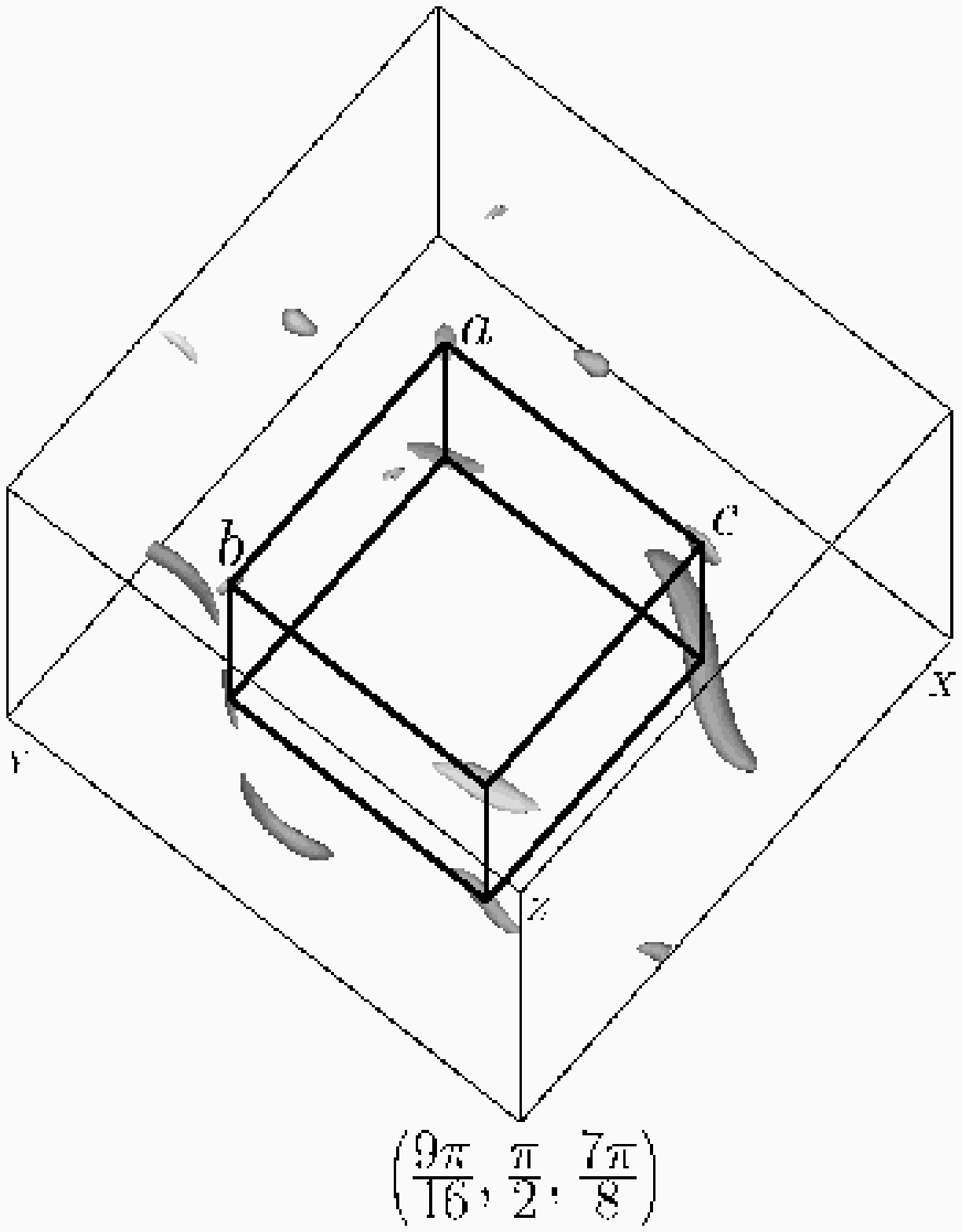,width=7cm,clip=}}
\vspace{-1truemm}
\centerline{\psfig{file=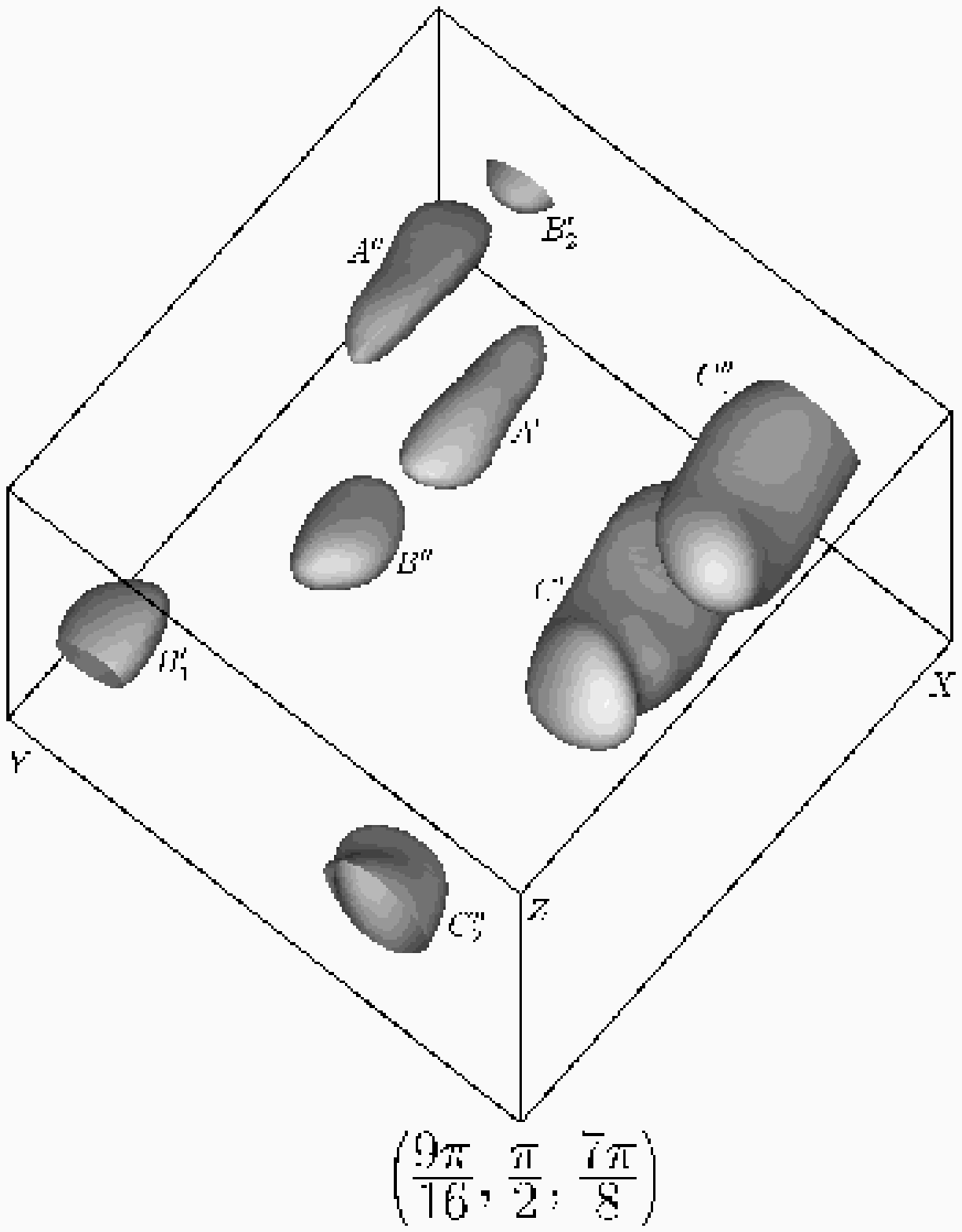,width=7cm,clip=}
\psfig{file=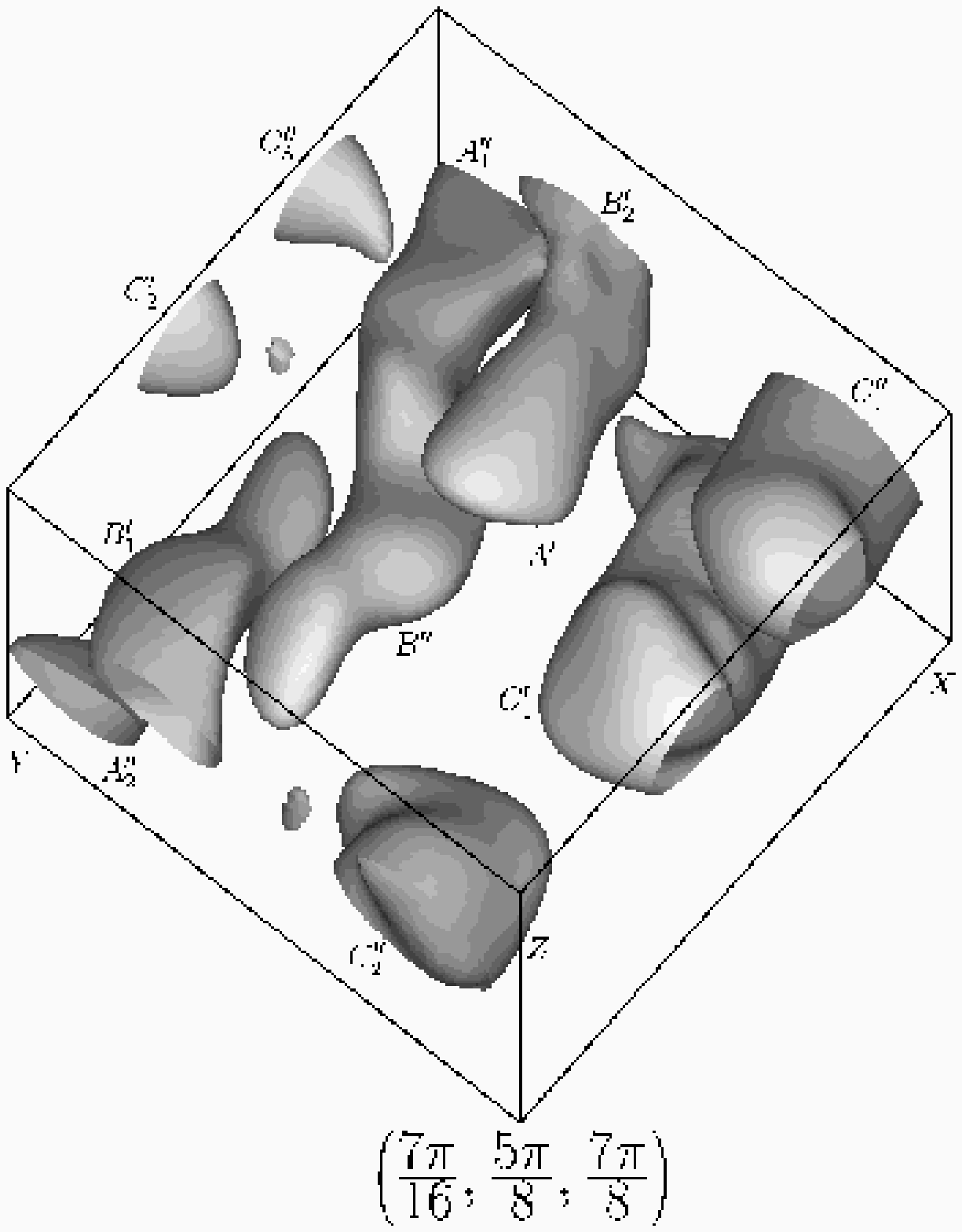,width=7cm,clip=}}
\else
\drawing 100 10 {eigenmode sample 2 (exponential spectrum)}
\fi
\caption{Second sample of velocity field  and
corresponding magnetic eigenfunction shown as isosurfaces. Upper:
$|{\bf v}|^2$ at 0.2\% of its maximum; lower left: $|{\bf H}|^2$ at
20\% of its maximum; lower right: $|{\bf H}|^2$ at 8\% of its
maximum. Otherwise as in
\xfg{f:sample1}.}
\label{f:sample2}
\end{figure}

Examples of magnetic eigenmodes for $\eta=0.1$ are shown in
Figures~\ref{f:sample1} and
\ref{f:sample2}. They correspond to two samples having exponential
velocity spectra. The upper of each figure shows regions where $|{\bf
v}|^2$ is less than 0.2\% of its maximum, hence revealing the
immediate neighborhood of stagnation points.  Note that in consequence
of parity invariance about the origin and of $2\pi$-periodicity,
within a basic $2\pi$-periodicity cube there are in total eight points
about which the flow is parity-invariant.  These stagnation points are
located at the vertices of a cube with side $\pi$ and they form a uniform
$\pi$-periodic grid in space.  The magnetic structures seen in the
lower left and right of each figure are clearly located near some of
those stagnation points. These structures are magnetic flux ropes of
the kind considered by Galloway and Zheligovsky (1994). Note that only
the fluctuating part of the magnetic eigenmodes are shown (both the
periodicity-box averaged magnetic field $\la{\bf h}_0\ra$
and the overall slow spatial
factor $e^{i\epsilon {\bf q}\cdot{\bf x}}$ are omitted). In
\xfg{f:sample2} $\la{\bf h}_0\ra$ is very
small (about two orders of magnitude smaller than the fluctuating
field), whereas for the case of \xfg{f:sample1} it is of the same
order of magnitude as the fluctuating field.

\begin{figure}
\iffigs
\centerline{\psfig{file=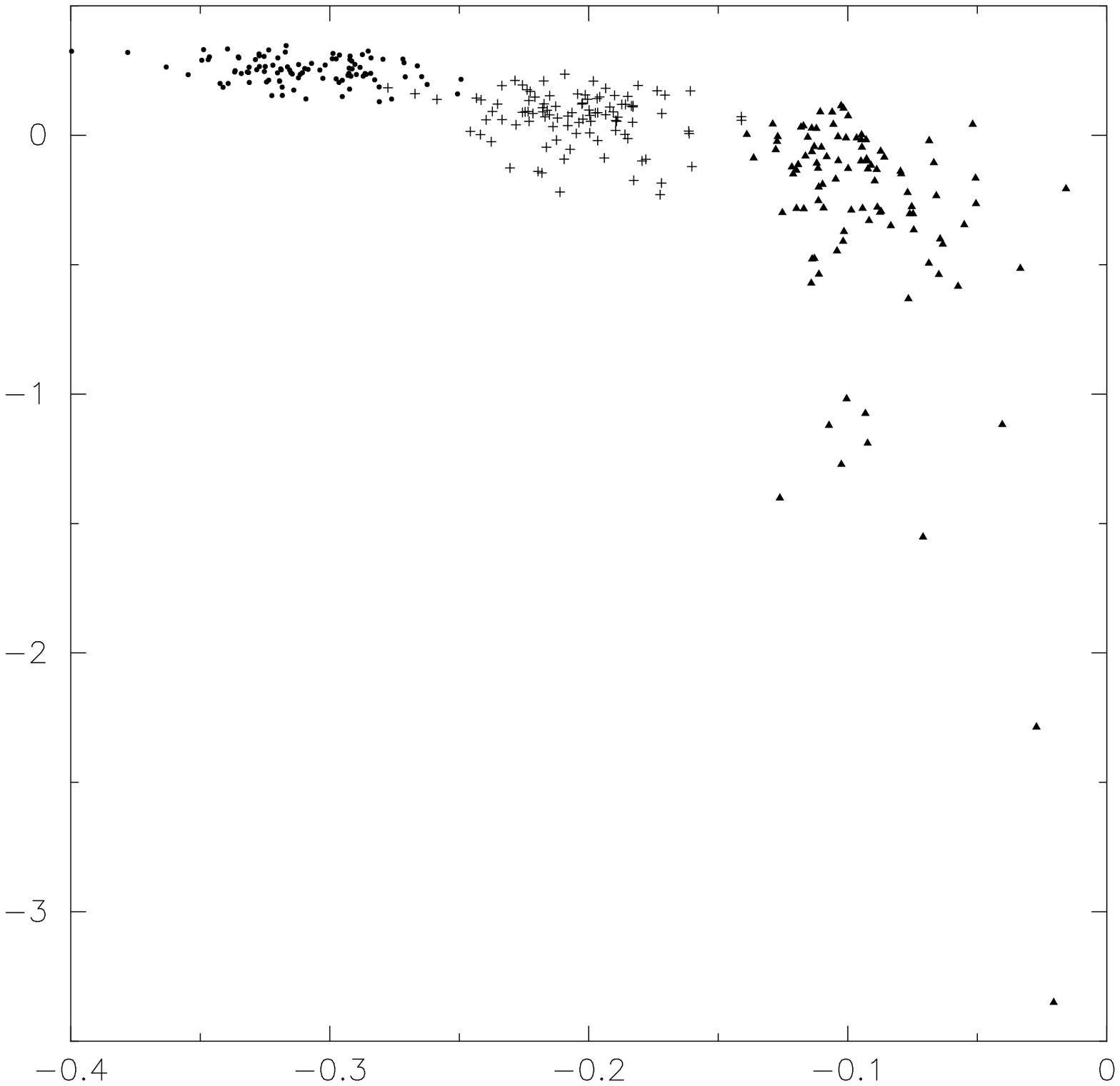,width=14cm,clip=}}
\else
\drawing 100 10 {eddy diffusivity vs growth rate (exponential spectrum)}
\fi
\caption{
Minimum eddy diffusivity $\eta_{\rm eddy}$ (vertical axis) versus the
growth rate of the corresponding dominant small-scale
magnetic modes (horizontal axis) for $\eta=0.3$ (dots), 0.2 (pluses)
and 0.1 (triangles) and 100 different flows with the exponential spectrum.}
\label{f:dotsntrianglesE}
\end{figure}

\begin{figure}
\iffigs
\centerline{\psfig{file=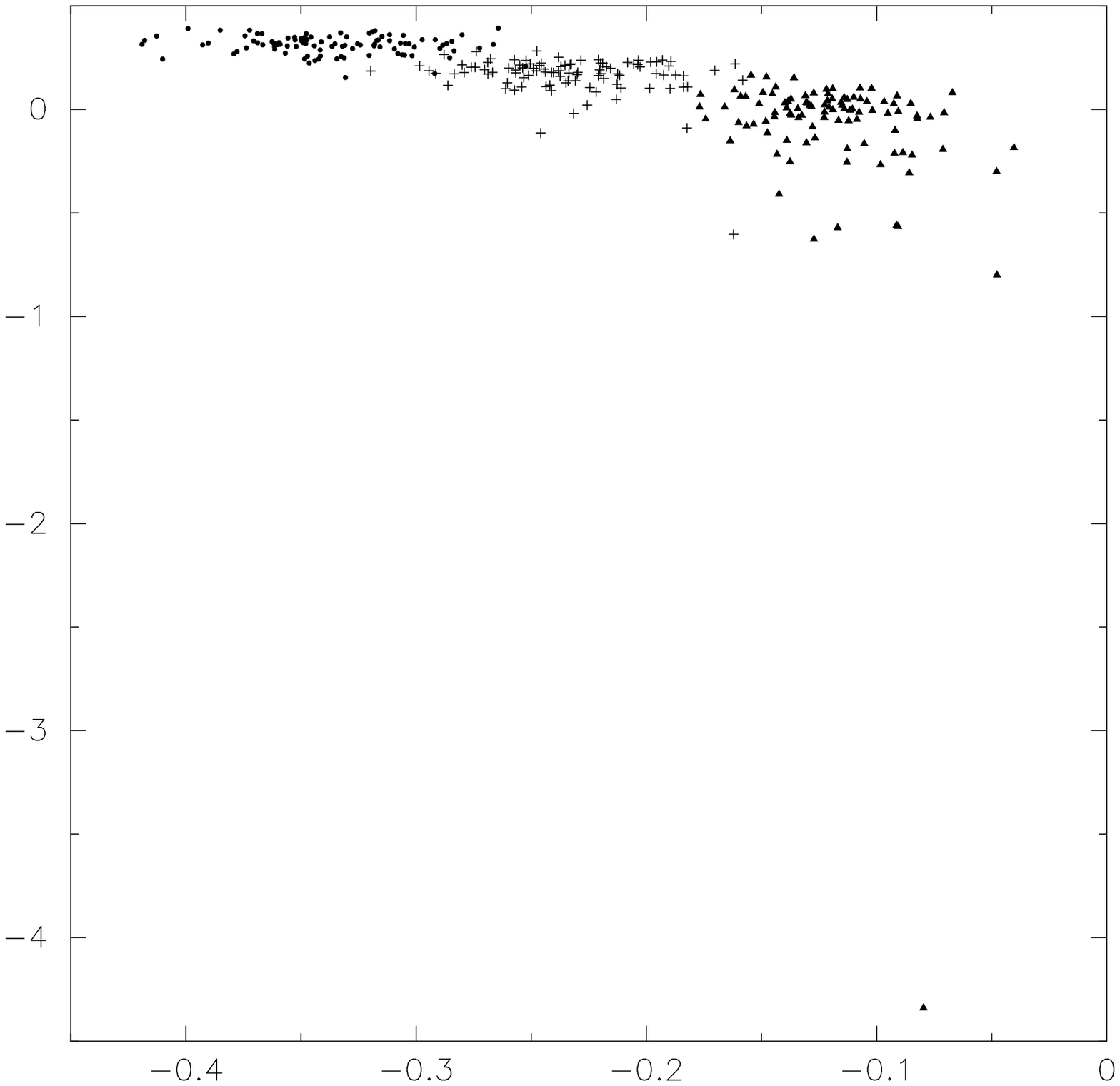,width=14cm,clip=}}
\else
\drawing 100 10 {eddy diffusivity vs growth rate (hyperbolic spectrum)}
\fi
\caption{
Same as \xfg{f:dotsntrianglesE} but with the hyperbolic spectrum.}
\label{f:dotsntrianglesH}
\end{figure}

A physically important question is: which instability arises first,
when magnetic molecular diffusivity is decreased, a large-scale or a
small-scale one? Similarly to what was done by Gama\et\
(1994), for each considered pair of a flow and a molecular diffusivity, it
is checked that small-scale magnetic fields with a vanishing mean are
stable, i.e. have negative growth rate (see
Figures~\ref{f:dotsntrianglesE} and~\ref{f:dotsntrianglesH}).

\begin{figure}
\iffigs
\centerline{\psfig{file=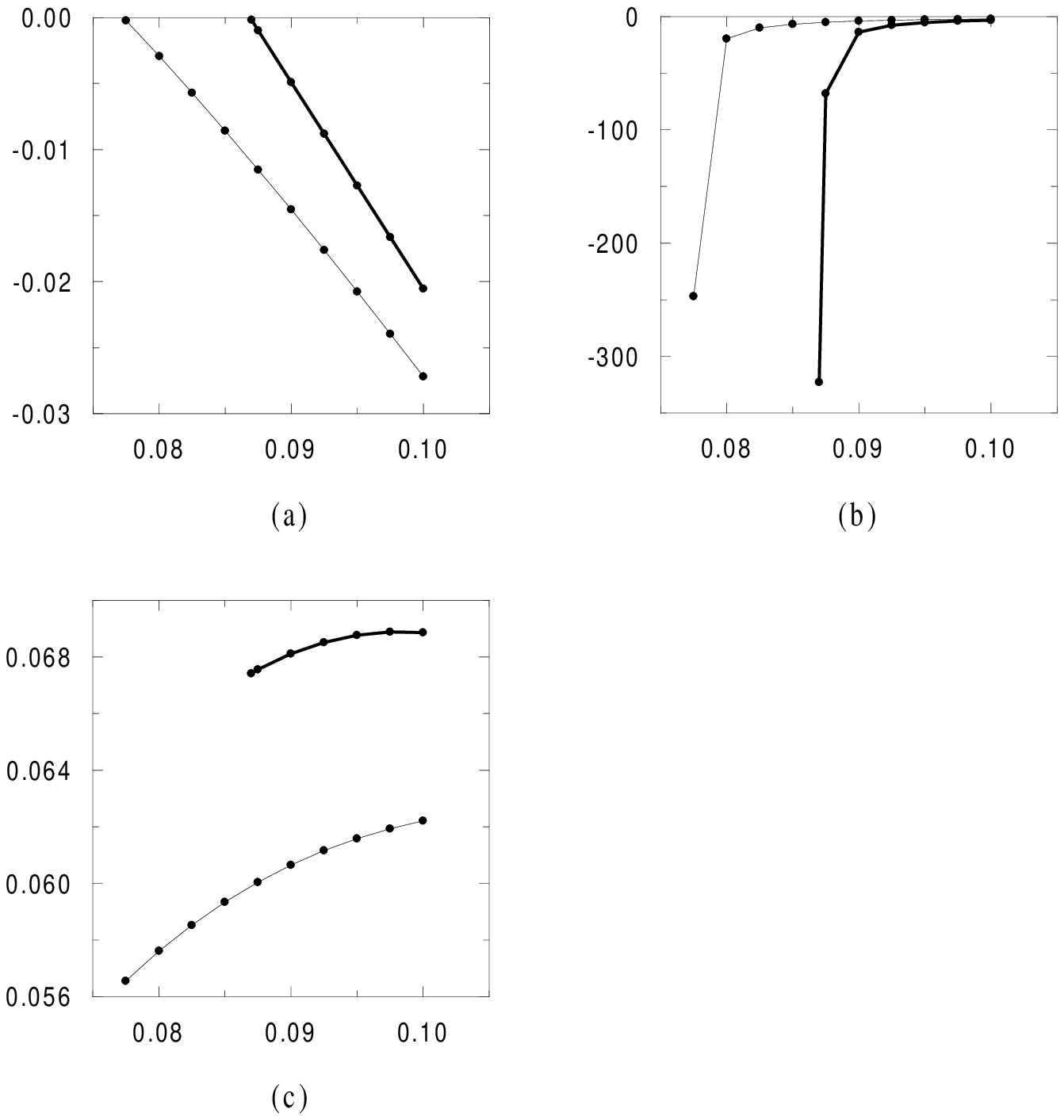,width=14cm,clip=}}
\else
\drawing 100 10 {infinitely low eddy diffusivity}
\fi
\caption{
Maximum growth rates $\zeta_1$ of small-scale magnetic modes (a),
minimum magnetic eddy diffusivity $\eta_{\rm eddy}$ (b) and product
$\zeta_1\eta_{\rm eddy}$ (c) all plotted versus molecular diffusivity
$\eta$. Line width codes two flow realizations, corresponding to the
two right-lower corner outliers in \xfg{f:dotsntrianglesE}.}
\label{f:lowdiffusivity}
\end{figure}

Several ``outliers'' with relatively large negative eddy
diffusivities can be seen on the histograms shown on \xfg{f:eddyhisto}
for $\eta=0.1$~. This can be explained as follows. We see on
Figures~\ref{f:dotsntrianglesE} and~\ref{f:dotsntrianglesH} that in the
case of outliers the growth rate of {\em small-scale} magnetic
modes (plotted horizontally) is negative, but close to zero.
Furthermore, we have checked that the corresponding eigenvalue is
real. In other words, we are close to a non-oscillatory
small-scale instability threshold. Now, we observe that the expression
of the eddy diffusivity, derived in \xsc{s:expansion} and
\xap{a:expansion}, involves the inverse of the small-scale magnetic induction
operator. It is thus not surprising that large eddy diffusivities can be
obtained. Depending on whether a certain inequality is satisfied at
the threshold, this eddy diffusivity can be either large negative or
large positive (see details in
\xap{a:loweddydiff}). We have checked numerically that the
outlier eddy diffusivities are approximately inversely proportional to
the small dominant eigenvalue (see \xfg{f:lowdiffusivity}).

Note, that if we are exactly at the threshold, our two-scale
expansion breaks down. A modified expansion can be carried out which
takes into account the enlargement of the kernel of the magnetic
induction operator $\cal L$. In the presence of a large-scale perturbation
this can lead to an eigenvalue $O(\epsilon)$ rather than the
$O(\epsilon^2)$-dependence obtained in the general parity-invariant
case. In other words, the large-scale dynamics can be governed
by a first-order differential operator.

\section{Generation of magnetic field with moderate scale separation}
\label{s:moderatescale}

For intermediate values of $\epsilon$, corresponding to moderate scale
separation, direct evaluation of the most unstable magnetic modes is desirable.
Computations are simplified by the fact that eigenmodes have the form
\rf{Hform}, all the dependence on the slow variable $\bf y=\epsilon x$
being absorbed into an exponential factor $e^{i\epsilon\bf q\cdot x}$.
Since the flow is assumed to depend only on the fast
variable $\bf x$, the eigenvalue problem is reduced to a
small-scale one for a {\em modified operator\/} by changing
$\nabla\to\nabla_{\bf x}+i\epsilon\bf q$ in the magnetic induction operator
\rf{eigeneq} and in the solenoidality condition \rf{Hsolenoidality}.

To find the most unstable magnetic mode one must, in
principle, allow for arbitrary orientations of the vector $\bf q$.
Throughout this section we assume only ``binary'' $\bf q$'s such that
each component takes only the values 0 or 1, so that $\epsilon^{-1}$ may
be regarded as the maximum scale-separation factor over cartesian directions.

Since our simulations are computationally demanding, we choose
for each flow only one $\bf q$-direction as follows.
For a given flow and for the molecular diffusivity
$\eta=0.1$, we obtain the most unstable $\bf q$-directions {\em in
the asymptotic regime\/} $\epsilon\to0$ from the results of
\xsc{s:abundance}. It turns out that for the cases reported in
this section, these directions are approximately parallel to binary vectors
with one or two non-zero components. These binary vectors
are then employed when $\epsilon$ is finite for all values of $\eta$ studied.

\begin{figure}
\iffigs
\centerline{\psfig{file=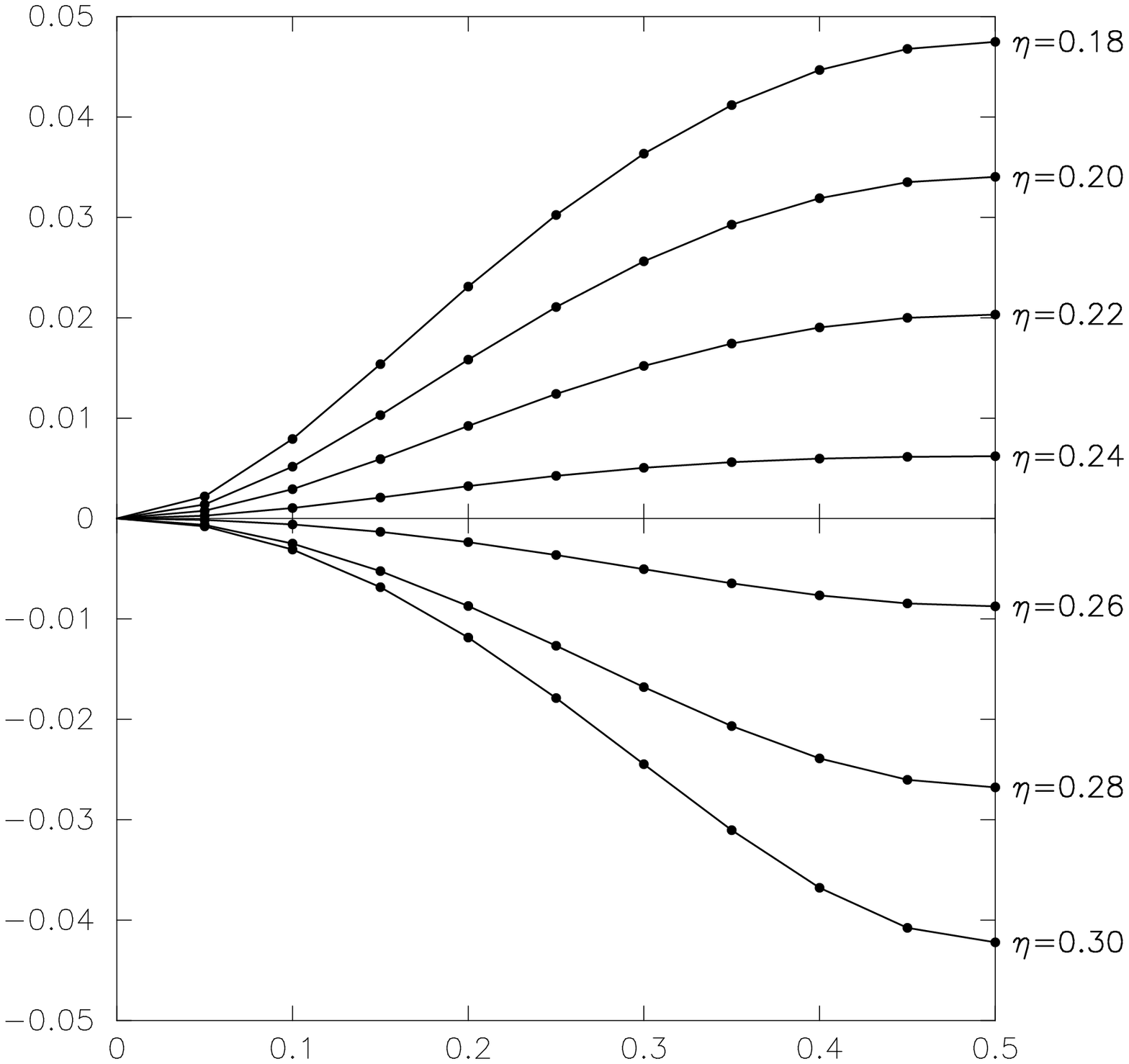,width=12cm,clip=}}
\else
\drawing 100 10 {growth rate vs scale ratio (a)}
\fi
\caption{
Maximum growth rates of magnetic modes (vertical axis) versus the
scale ratio $\epsilon$ (horizontal axis) for different molecular
diffusivities $\eta$ and one of four samples of a flow with the hyperbolic
spectrum.}
\label{f:growthvsepsilonA}
\end{figure}

\begin{figure}
\iffigs
\centerline{\psfig{file=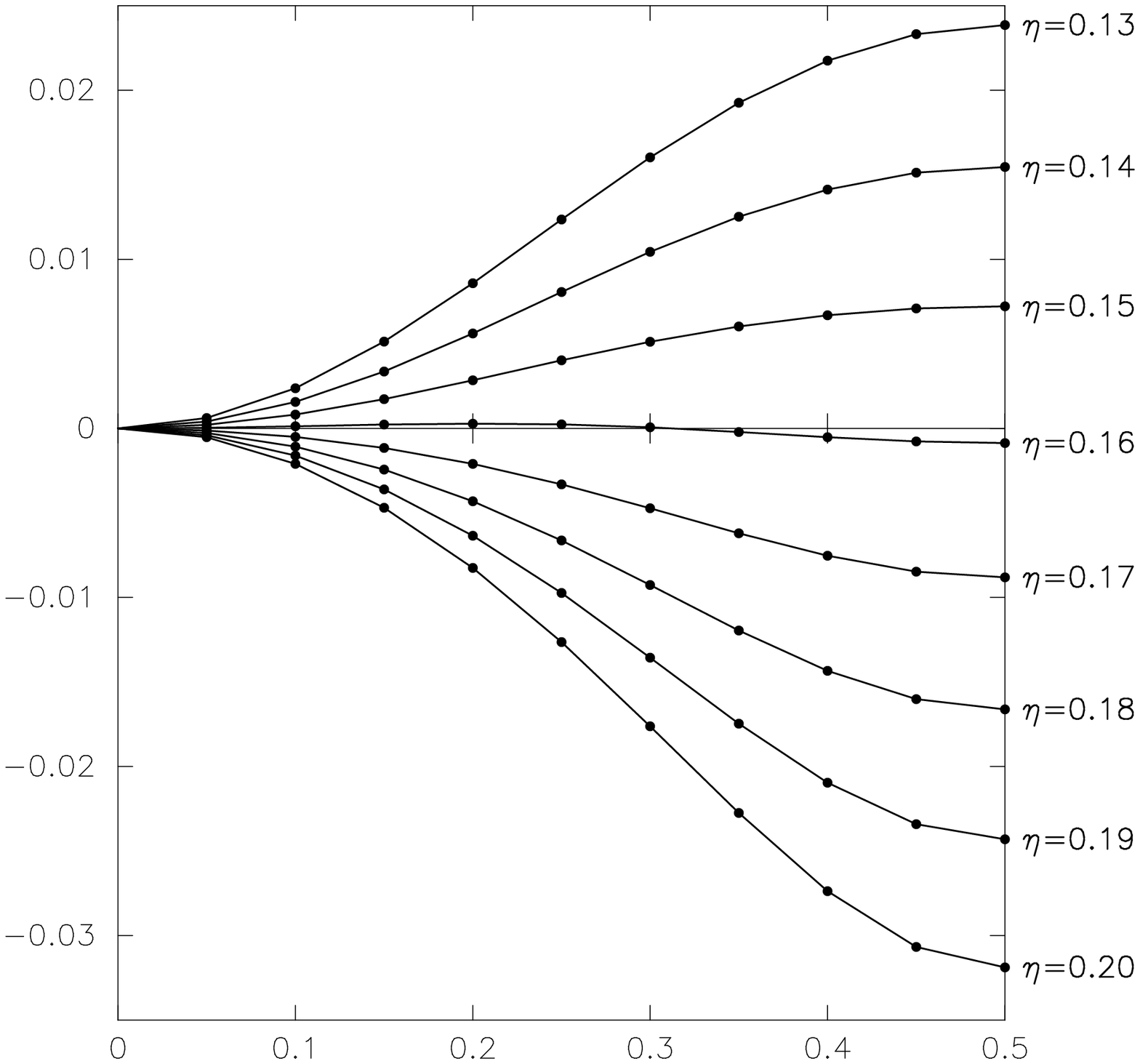,width=12cm,clip=}}
\else
\drawing 100 10 {growth rate vs scale ratio (b)}
\fi
\caption{
Same as \xfg{f:growthvsepsilonA}, another flow sample.}
\label{f:growthvsepsilonB}
\end{figure}

\begin{figure}
\iffigs
\centerline{\psfig{file=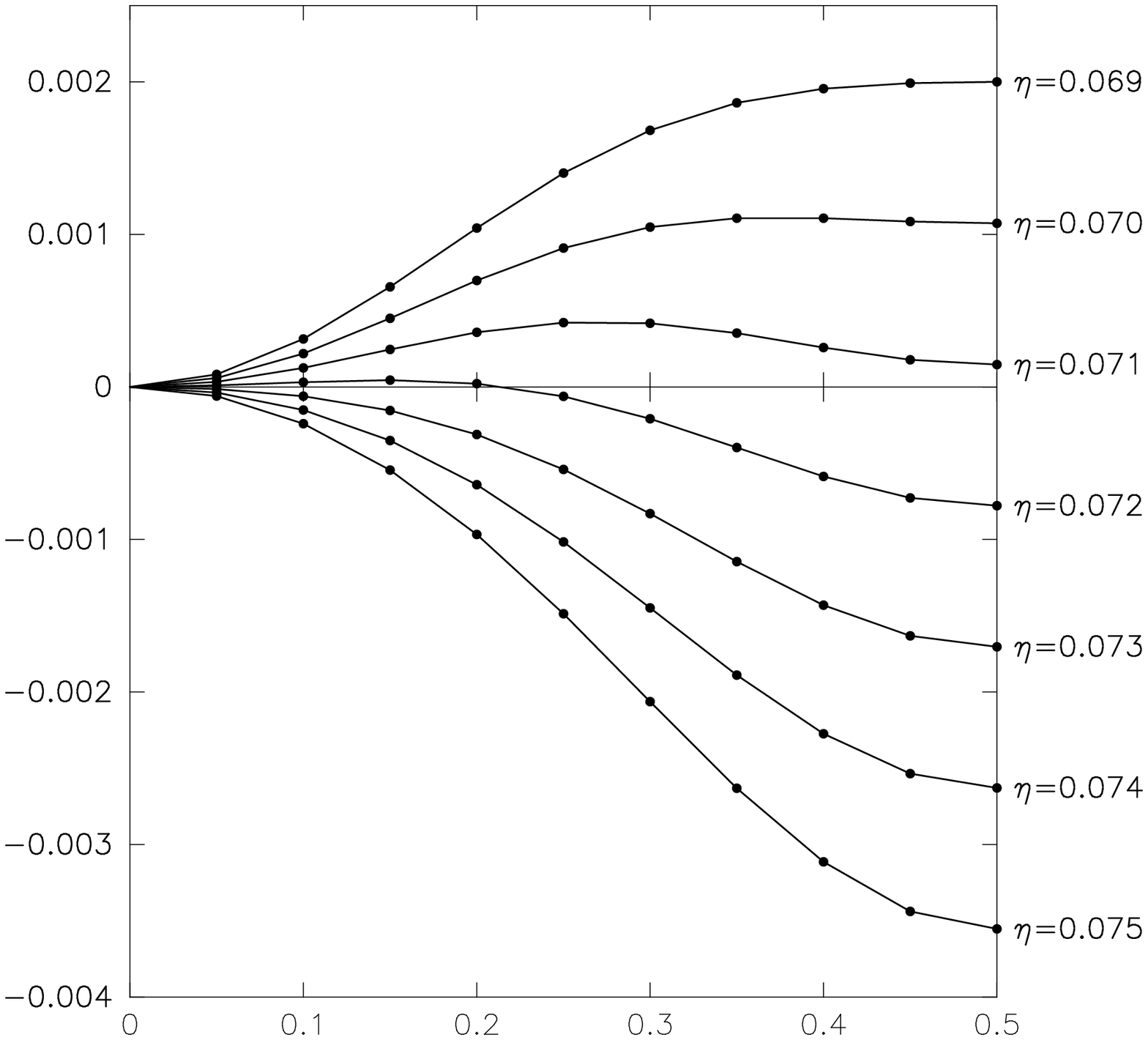,width=12cm,clip=}}
\else
\drawing 100 10 {growth rate vs scale ratio III}
\fi
\caption{
Same as \xfg{f:growthvsepsilonA}, yet another flow sample.}
\label{f:growthvsepsilonC}
\end{figure}

\begin{figure}
\iffigs
\centerline{\psfig{file=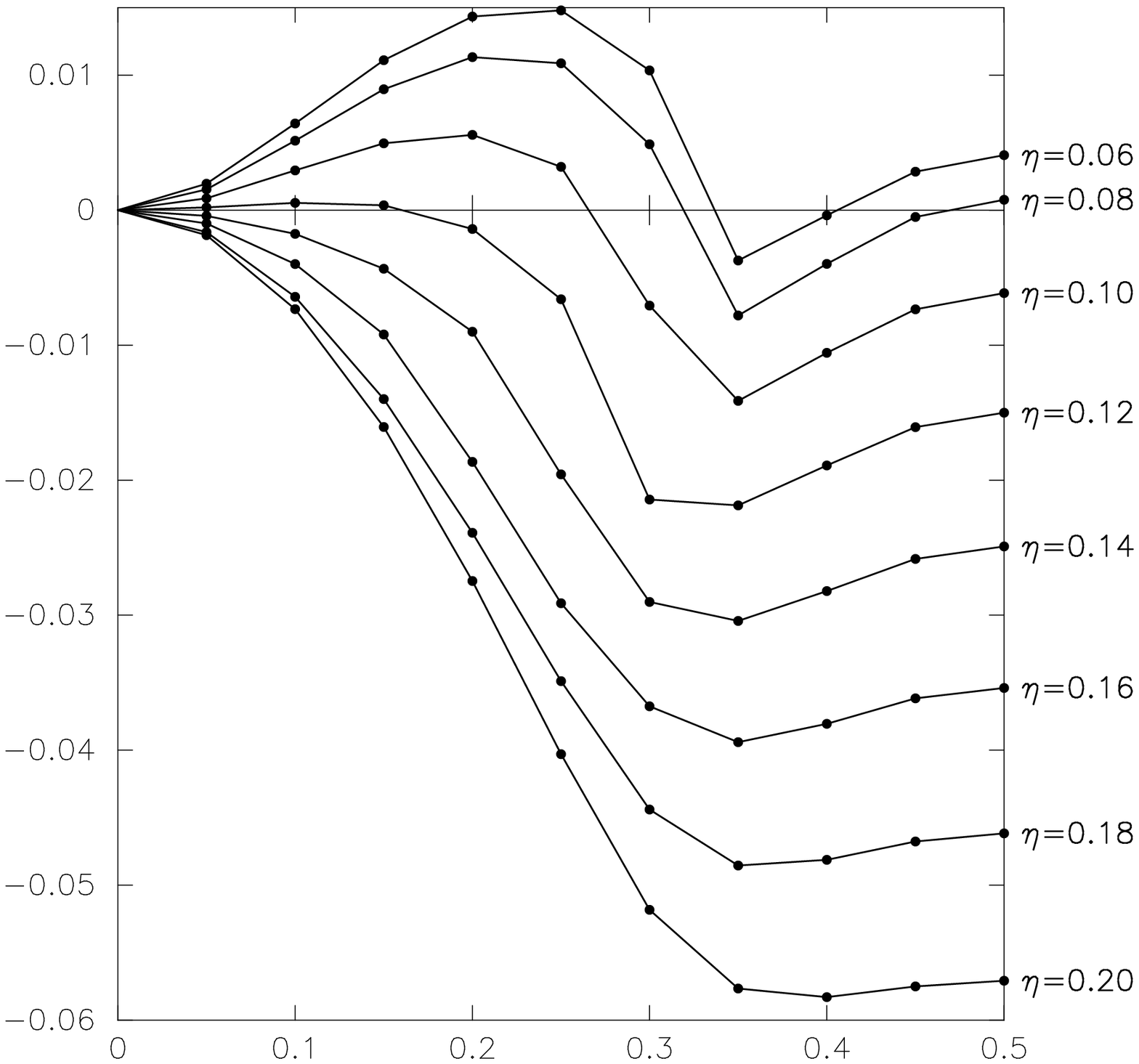,width=12cm,clip=}}
\else
\drawing 100 10 {growth rate vs scale ratio IV}
\fi
\caption{
Same as \xfg{f:growthvsepsilonA}, the last flow sample.}
\label{f:growthvsepsilonD}
\end{figure}

Figures~\ref{f:growthvsepsilonA}-\ref{f:growthvsepsilonD} show the
dominant growth rates for four different flows selected out of the
ensemble of flows with hyperbolic spectrum, which sustain negative eddy diffusivity.
Actually, when $\bf q$ has integer components, it is only necessary to
make computations for $0\le\epsilon\le1/2$. Indeed, first note, that the
modified magnetic induction operators for two values of $\epsilon$
differing by 1 have the same spectrum and their eigenfunctions differ by
a factor $e^{i\bf qx}$. Second, by parity invariance of the velocity
\rf{vparityinv}, for two opposite values of $\epsilon$ the spectra
are also the same and eigenfunctions are related by parity ${\bf
H}({\bf x})\to{\bf H}(-{\bf x})$. (From this one easily deduces that
the spectrum of the modified operator is complex conjugate,
eigenvalues $\lambda$ and $\overline{\lambda}$ being associated to
eigenfunctions ${\bf H}({\bf x})$ and $\overline{\bf H}(-{\bf x})$,
respectively.) Hence, outside the interval $0\le\epsilon\le1/2$ the
plots can be continued by symmetry about the vertical line
$\epsilon=1/2$ and by 1-periodicity.

The results reported here are obtained with the resolution of $64^3$
Fourier harmonics. For the ``worst'' case, $\eta=0.07$, they differ
only in the third significant digit from those obtained with the
resolution $32^3$. Note, that energy spectra of the computed magnetic
modes decrease by at least 4 orders of magnitude for $\eta=0.13$ and
by at least 5 for $\eta=0.3$~, indicating that the employed resolution
is adequate. (For $\eta=0.07$ the decrease is only by 3 orders of
magnitude, which may seem marginal. However, it is well-known that
eigenvalues are less sensitive to insufficient resolution, than the
fine structure of eigenmodes.) It is clear from
Figures~\ref{f:growthvsepsilonA}-\ref{f:growthvsepsilonD}
that for all 
instances where the eddy diffusivity is negative (i.e.\ the graph go to
zero by negative values as  $\epsilon \to 0$) there is also an instability
for any finite $\epsilon$. The converse is, however, not always true;
for example, for the last flow (Figure~\ref{f:growthvsepsilonD})
and $\eta=0.10$ there is a dynamo for $0.26<\epsilon \le 0.5$, which 
disappears below $\eta=0.26$ (the eddy diffusivity is positive).

Furthermore, inspection of
Figures~\ref{f:growthvsepsilonB}-\ref{f:growthvsepsilonD} reveals that
there can be one or more windows in scale separation, where the dynamo
(positive growth rate) is present, whereas in
Figures~\ref{f:growthvsepsilonA} the dynamo appears to be present for
all scale separations, provided the eddy diffusivity is negative. In
\xfg{f:growthvsepsilonD} the two windows belong to
two different analytical branches of eigenvalues and eigenmodes, with
the transition occurring near $\epsilon=0.3$~.

Note, that all the growth rates vanish (quadratically) when $\epsilon\to0$.
The reason is that the small-$\epsilon$ eigenvalues, obtained
by the two-scale expansion of \xsc{s:expansion}, are analytic continuations
of the zero eigenvalue in the small-scale problem. The presence of the
eigenvalue zero is due to the structure of the magnetic induction operator,
which is the curl of a first-order operator. (Hence, its adjoint,
acting also in the space of divergenceless functions
$$
{\cal L}^*: \,\,\,{\bf H}\mapsto\eta\nabla_{\bf x}^2{\bf H}
+(\nabla_{\bf x}\times{\bf H})\times{\bf v}-\nabla_{\bf x}p
$$
has constant vectors in its kernel.) Conversely, it can be demonstrated
by spatial averaging of ${\cal L}{\bf H}=\lambda\bf H$, that for a mode with
a non-vanishing mean the associated eigenvalue equals 0. (Note, that in
\xsc{s:abundance} we restricted the operator to functions of
vanishing spatial mean and thus the growth rate zero was absent from
Figures~\ref{f:dotsntrianglesE} and \ref{f:dotsntrianglesH}.)

\begin{figure}
\iffigs
\centerline{\psfig{file=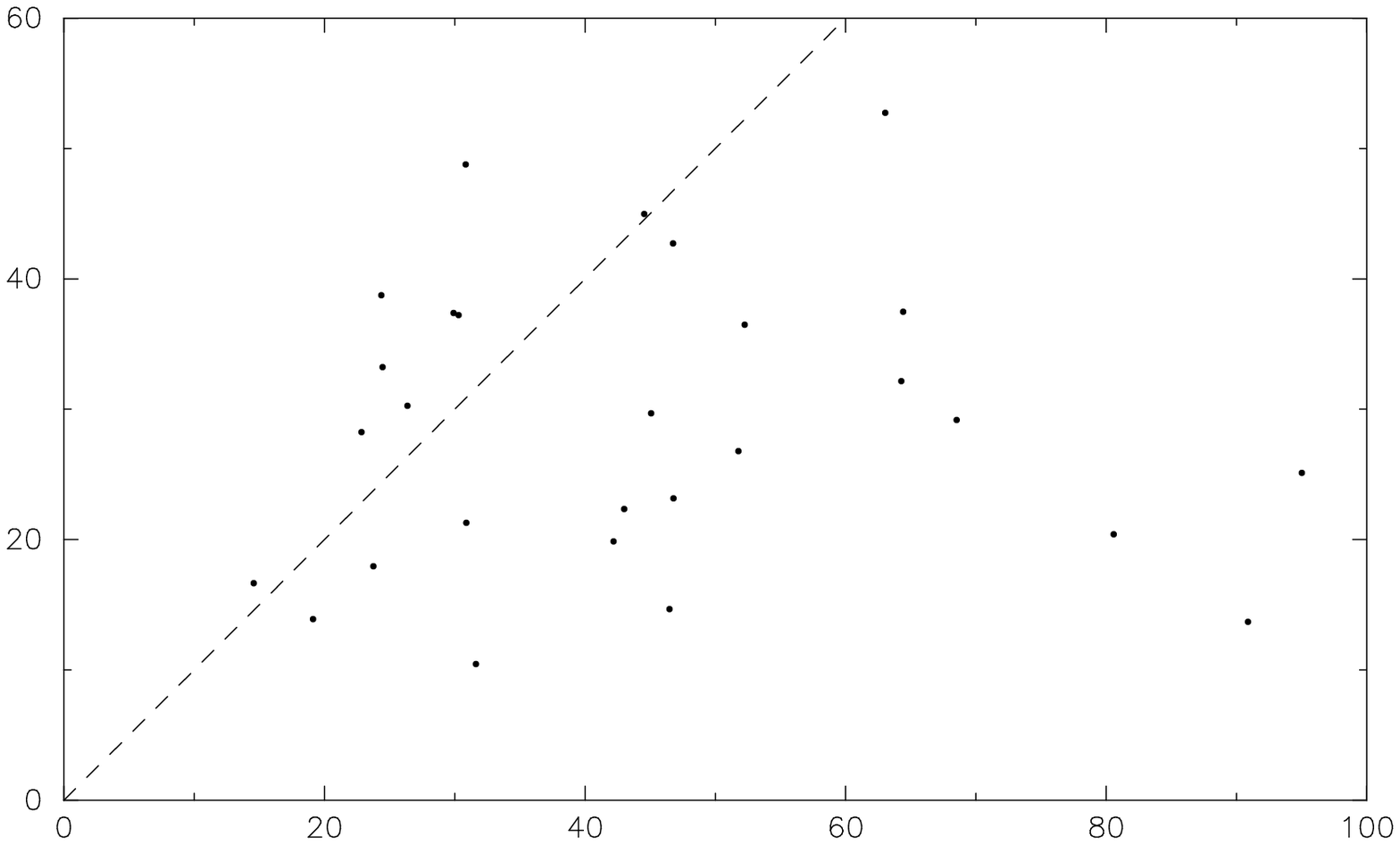,width=14cm,clip=}}
\else
\drawing 100 10 {critical Rm's: small-scale vs double period}
\fi
\caption{
Critical magnetic Reynolds numbers for the onset of instability of small-scale
magnetic modes (horizontal axis) versus critical magnetic Reynolds numbers
for magnetic modes of twice the spatial period (vertical axis).
28 flows with the hyperbolic spectrum are shown.
The dashed line corresponds to equal Reynolds numbers.}
\label{f:criticalRm}
\vspace*{.5in}
\end{figure}

We now turn to the case, where there is a finite scale separation of a
factor two ($\epsilon=1/2$). The question we address is: how does a
mere doubling of the spatial period of magnetic modes (relative to
that of the basic flow) affect generation? More precisely: to
lower the critical magnetic Reynolds number, does it help to
introduce such a scale separation? We systematically compare critical
numbers when $\epsilon=1$ (no scale separation) and $\epsilon=1/2$ for
an ensemble of flows with hyperbolic energy spectrum. We
define our magnetic Reynolds numbers $R_m=LV/\eta$ in terms of the
{\em largest\/} scale available $L=\epsilon^{-1}$ (the inverse of the
minimum wavenumber of the magnetic field) and of the r.m.s. velocity
$V$. In computations $V=1$ and hence $R_m=(\epsilon\eta)^{-1}$.

When $\epsilon=1/2$ we
face again the problem, mentioned earlier, of choosing the vectors $\bf q$.
For each flow we now search for the critical $R_m$, allowing all 7
possible binary ${\bf q}\ne0$. More precisely, we find an
approximate (minimum) critical $R_m$ and the corresponding $\bf q$ by running
our code with $32^3$ Fourier harmonics, and then refine the critical $R_m$
with $64^3$ Fourier harmonics for the obtained $\bf q$.

The secant method is used to find stationary modes. Iterations arere
terminated when the real part of the dominant eigenvalue was below
$10^{-5}$ in magnitude. This usually requires 4-5 iterations.

As far as the accuracy of the calculation is concerned, note
that in all $\epsilon=1/2$ runs with $R_m>25$ the energy spectrum
decreases by at least 4 orders of magnitude, an indication that
$64^3$ harmonics provide sufficient resolution. In the most demanding
runs, with $\epsilon=1$ and $R_m>50$, the energy spectra decrease
only by 2--3 orders of magnitude but, as pointed out before,
eigenvalues are computed nevertheless quite accurately (critical $R_m$'s,
computed with the $32^3$ and $64^3$ resolutions, differ in the third
significant digit).

\xfg{f:criticalRm} shows all the critical $R_m$'s
calculated for 28 cases. Two cases not shown have critical $R_m=30.7$
and $25.3$ when $\epsilon=1/2$, but do not display dynamo action for
any $R_m\le 100$ when $\epsilon=1$. The results demonstrate that scale
separation lowers the critical $R_m$ in all but 9 cases out of the 30
studied.  Note that Nore\et\ (1997) have studied a flow (a variant of
Taylor--Green) for which a scale separation $\epsilon=1/2$ seems to
lower significantly the critical magnetic Reynolds number, as is the
case for the helical ABC flows studied by Galanti\et\ (1992), Archontis
(2000) and Dorch (2000).  However the Nore\et\ (1997) definition of the 
magnetic Reynolds number uses the scale of the flow and not that
of the magnetic field. If instead our definition is used, their flow appears 
to be just a borderline case. 

\section{Conclusion}
\label{s:conclusion}

Three main results are presented in this work.
First, in \xsc{s:abundance}, we demonstrated that parity-invariant flows
possessing a negative magnetic eddy diffusivity are quite common.
Specifically, we generated two random ensembles of flows;
both had time-independent randomly chosen Fourier components, one with
an exponentially decreasing energy spectrum, the other -- with a power-law
decrease (referred to as ``hyperbolic'' because the exponent is
-1). One hundred independent selections were made in each case and
three different molecular magnetic diffusivities were used, all
corresponding to small-scale magnetic Reynolds numbers not exceeding
ten. None of the 600 cases investigated was a small-scale dynamo.
For the lowest diffusivity considered, 86\% of the ``exponential'' spectrum
flows and 53\% of the ``hyperbolic'' spectrum flows produced a negative eddy
diffusivity and,
hence, can act as a large-scale dynamo (for more details, see Table~1).

Second, several instances with particularly negative eddy diffusivities were
encountered. The theory of these ``outliers'' (given in details in
\xap{a:loweddydiff}) relates them to a near-critical state,
where an eigenvalue of the small-scale dynamo problem is about to pass through
zero (there is also an additional inequality ensuring that a large
{\em negative\/} value is obtained).

Finally, we presented numerical evidence that, for a parity-invariant
flow, even a moderate scale separation between the flow and the
magnetic field is generally beneficial for magnetic field
generation. Specifically, we showed that the critical magnetic
Reynolds number based on the largest scale present (that of the
magnetic field) is often lowered when the scale of twice the spatial period of
the velocity is introduced (70\% of the instances studied had this property;
see \xfg{f:criticalRm}). For a few flows we also studied the
dependence on the scale-separation factor, when it is varied from a
1:2 ratio to a 1:20 ratio. One instance (shown in
\xfg{f:growthvsepsilonD}) has a window lacking dynamo action
between low and large values, where it is present. 

For helical flow, such as ABC, the lowering of the critical magnetic
Reynolds number, when a moderate separation of scale is assumed, is
understood in terms of symmetry breaking, stagnation points and other
geometrical features (Archontis 2000). It would be of interest to find
which geometrical features are important for the lowering mechanism
when the flow is parity-invariant.

\noindent{\bf Acknowledgments}.

We are grateful to the authors of Lanotte\et\ (1999) for making
available to us the source of the code used to obtain the results
reported in their paper, to S.~Fauve, S.~Gama, A.~Lanotte, A.~Noullez,
Y.~Ponty and M.~Vergassola for useful discussions and to an anonymous
referee for several suggestions. Computational facilities were
provided by the program ``Simulations Interactives et Visualisation en
Astronomie et M\'ecanique (SIVAM)'' at the Observatoire de la C\^ote
d'Azur. Visits of O.M.~Podvigina and V.A.~Zheligovsky were supported
by the French Ministry of Education. This work was also supported by
the European Union under contract HPRN-CT-2000-00162.

\pagebreak
\section*{Appendices}
\appendix
\section{Formal asymptotic expansion
of magnetic modes and the associated eigenvalues}
\setcounter{equation}{0}
\renewcommand{\theequation}{\thesection.\arabic{equation}}
\label{a:expansion}

We construct here complete asymptotic expansions of magnetic eigenmodes and
of their eigenvalues for small values of the scale ratio $\epsilon$.
Assumptions concerning the flow are listed in \xsc{s:expansion}.

The kinematic dynamo problem reduces to the eigenvalue problem
\begin{equation}
{\cal M}{\bf H}=\lambda{\bf H}.
\label{Meigeneq}\end{equation}
Here ${\bf H}({\bf x})$ is a magnetic mode, $\lambda$ is its eigenvalue, and
$$
{\cal M}{\bf H}\equiv\eta\nabla^2{\bf H}+\nabla\times({\bf v}\times{\bf H}),
$$
acting in the space of solenoidal vector fields, is the magnetic
induction operator (it acts like $\cal L$, but in a larger
domain). The magnetic mode ${\bf H}({\bf x},{\bf y})$ is supposed to
be solenoidal \rf{Hsolenoidality} and $2\pi$-periodic in both fast and
slow variables. Solutions to the eigenproblem \rf{Meigeneq} are sought
in the form of asymptotic series $\bf H=\sum{\bf h}_n\epsilon^n$ and
$\lambda=\sum\lambda_n\epsilon^n$. Denote
$$
{\bf H}_n\equiv\la{\bf h}_n\ra;\quad{\bf G}_n\equiv\lb{\bf h}_n\rb,
$$
where $\la\cdot\ra$ and $\lb\cdot\rb$ refer
to the mean and fluctuating part (see \rf{defmeanosc}~), respectively.

The expansion is constructed under the assumption that zero is not an
eigenvalue of the operator $\cal L$ acting in the space of small-scale
periodic solenoidal vector fields with vanishing mean. (By ``small-scale''
we understand ``depending only on the fast variables''.)

The decomposition of the gradient
$\nabla=\nabla_{\bf x}+\epsilon\nabla_{\bf y}$, applied to the solenoidality
condition, expanded in powers of $\epsilon$ and separated into mean
and fluctuating part, gives
\begin{equation}
\nabla_{\bf y}\cdot{\bf H}_n=0,
\label{Hnsolenoidality}\end{equation}
\begin{equation}
\nabla_{\bf x}\cdot{\bf G}_n+\nabla_{\bf y}\cdot{\bf G}_{n-1}=0
\label{Gnsolenoidality}\end{equation}
for all $n\ge0$. (${\bf H}_n={\bf G}_n\equiv0$ is assumed for all $n<0$.)

Substituting the expansions and the decomposition of the gradient
into \rf{Meigeneq}, we obtain:
\begin{eqnarray}
&&\left.\sum\limits_{n=0}^\infty\right[{\cal L}{\bf G}_n+\eta\left(
2(\nabla_{\bf x}\cdot\nabla_{\bf y}){\bf G}_{n-1}
+\nabla^2_{\bf y}({\bf H}_{n-2}+{\bf G}_{n-2})\right)
+\nabla_{\bf x}\times({\bf v}\times{\bf H}_n)\nonumber\\
&&\left.+\,\,\nabla_{\bf y}\times({\bf v}\times({\bf H}_{n-1}+{\bf G}_{n-1}))
-\sum\limits_{m=0}^n\lambda_{n-m}({\bf H}_m+{\bf G}_m)\right]\epsilon^n=0.
\label{serieseqn}\end{eqnarray}
We proceed by successively equating to zero the mean and the fluctuating part
of each term of this series. In our context, averaging
is equivalent to obtaining the solvability condition for a partial differential
equation in fast variables.

$i$. Transform the leading $(n=0)$ term of \rf{serieseqn} using vector algebra
identities and \rf{vsolenoidality}:
\begin{equation}
{\cal L}{\bf G}_0+({\bf H}_0\cdot\nabla_{\bf x}){\bf v}=
\lambda_0({\bf H}_0+{\bf G}_0).
\label{eigeneq0}\end{equation}
Averaging of this equation yields $\lambda_0=0$. Then the solution of
\rf{eigeneq0} can be expressed as
\begin{equation}
{\bf G}_0=\sum_{k=1}^3{\bf S}_k({\bf x})H^k_0({\bf y}),
\label{G0}\end{equation}
where the vector fields ${\bf S}_k$ are the solutions of the first
auxiliary problem \rf{Seq}; the solutions exist since it is
assumed that the kernel of $\cal L$ is empty. (The first and second
auxiliary equations were stated in \xsc{s:expansion}.)

By taking the divergence of the first auxiliary equation \rf{Seq}, one
obtains $\nabla_{\bf x}\cdot{\bf S}_k=0$. Since the velocity is
parity-invariant, the domain of $\cal L$ is a direct sum of two invariant
subspaces, one of which is comprised of parity-invariant vector fields
(i.e.\ ${\bf f}({\bf x})=-{\bf f}(-{\bf x})$~), and the second -- of the
parity-antiinvariant ones (${\bf f}({\bf x})={\bf f}(-{\bf x})$~).
Thus, $\bf S$ is a parity-antiinvariant matrix:
\begin{equation}
{\bf S}_k({\bf x})={\bf S}_k(-{\bf x}).
\label{Snonparityinv}\end{equation}

$ii$. The $\epsilon^1$ term of \rf{serieseqn}, after the use of
\rf{Hnsolenoidality}, takes the form
\begin{eqnarray}
&&{\cal L}{\bf G}_1+2\eta(\nabla_{\bf x}\cdot\nabla_{\bf y}){\bf G}_0
+({\bf H}_1\cdot\nabla_{\bf x}){\bf v}+\nabla_{\bf y}\times({\bf v}
\times{\bf G}_0)\nonumber\\[0.8ex]
&&-({\bf v}\cdot\nabla_{\bf y}){\bf H}_0=\lambda_1({\bf H}_0+{\bf G}_0).
\label{eigeneq1}\end{eqnarray}
Substituting \rf{G0} in \rf{eigeneq1} and averaging, we obtain
\begin{equation}
\nabla_{\bf y}\times(\,\la{\bf v}\times{\bf S}\ra\,{\bf H}_0)=\lambda_1{\bf H}_0.
\label{eigeneq1mean}\end{equation}

So far, developments strictly followed those of Vishik (1987), but
here the point of divergence is reached. Since $\bf v$ is parity-invariant
\rf{vparityinv} and $\bf S$ is parity-antiinvariant \rf{Snonparityinv},
we have $\la{\bf v}\times{\bf S}\ra=0$, and therefore \rf{eigeneq1mean}
implies $\lambda_1=0$. Upon inserting \rf{G0} into the fluctuating part of
\rf{eigeneq1}, we obtain
\begin{equation}
{\cal L}{\bf G}_1=-({\bf H}_1\cdot\nabla_{\bf x}){\bf v}
+\sum_{k=1}^3\sum_{m=1}^3\left(-2\eta{\partial{\bf S}_k\over\partial x_m}
+v^m({\bf S}_k+{\bf e}_k)-{\bf v}S^m_k\right)
{\partial{\bf H}^k_0\over\partial y_m},
\label{eigeneq1osc}\end{equation}
from which one finds
\begin{equation}
{\bf G}_1=\sum_{k=1}^3{\bf S}_k({\bf x})H^k_1({\bf y})+
\sum_{k=1}^3\sum_{m=1}^3{\bf\Gamma}_{mk}({\bf x})
{\partial H^k_0\over\partial y_m}({\bf y}).
\label{G1}\end{equation}
Here, the nine vector fields ${\bf\Gamma}_{mk}$ are solutions to the
second auxiliary equation \rf{Gammaeq}. Note
${\bf\Gamma}_{mk}({\bf x})=-{\bf\Gamma}_{mk}(-{\bf x})$ by the symmetry
properties of $\bf v$, ${\bf S}_k$ and of the operator $\cal L$.
By taking divergence of the second auxiliary equation one finds
$\nabla_{\bf x}\cdot{\bf\Gamma}_{mk}+{\bf S}^m_k=0$. Hence, for $n=1$
\rf{Gnsolenoidality} is automatically satisfied.

$iii$. The $\epsilon^2$ term of \rf{serieseqn} takes the form
\begin{eqnarray}
&&{\cal L}{\bf G}_2+\eta\left(2(\nabla_{\bf x}\cdot\nabla_{\bf y}){\bf G}_1
+\nabla^2_{\bf y}({\bf H}_0+{\bf G}_0)\right)+({\bf H}_2
\cdot\nabla_{\bf x}){\bf v}\nonumber\\[0.8ex]
&&+\,\,\nabla_{\bf y}\times({\bf v}\times{\bf G}_1)
-({\bf v}\cdot\nabla_{\bf y}){\bf H}_1=\lambda_2({\bf H}_0+{\bf G}_0).
\label{eigeneq2}\end{eqnarray}
Upon substitution of \rf{G1}, averaging produces the second-order equation
\begin{equation}
\overline{\cal M}{\bf H}_0\equiv\eta\nabla^2_{\bf y}{\bf H}_0+
\nabla_{\bf y}\times
\sum_{k=1}^3\sum_{m=1}^3\la{\bf v}\times{\bf\Gamma}_{mk}
\ra{\partial H^k_0\over\partial y_m}=\lambda_2{\bf H}_0,
\label{wasnotnumbered}
\end{equation}
which defines the eddy diffusivity operator. So, the leading terms of
the expansions of the mean magnetic field, ${\bf H}_0$, and of the
associated eigenvalue, $\lambda_2$, satisfy the eigenvalue equation
for this operator. Its eigenvectors are Fourier harmonics:
${\bf H}_0={\bf h}e^{i\bf q\cdot y}$, where $\bf h$ and $\bf q$ are constant
vectors satisfying \rf{3deigeneq} and \rf{3dsolenoidality}.
Note that the fluctuating part of the leading term of the eigenmode,
${\bf G}_0$, is now completely determined by \rf{G0}.

At this point our construction goes beyond that of Lanotte\et\ (1999)
and, in the spirit of Vishik (1987), we show how to extend the analysis
to higher orders in $\epsilon$.

The fluctuating part of \rf{eigeneq2} reads
$${\cal L}{\bf G}_2=-\eta\left(2\sum_{k=1}^3\sum_{m=1}^3
\left({\partial{\bf S}_k\over\partial x_m}{\partial H^k_1\over\partial y_m}
+\sum_{l=1}^3{\partial{\bf\Gamma}_{mk}\over\partial x_l}
{\partial^2H^k_0\over\partial y_m\partial y_l}\right)
+\nabla^2_{\bf y}{\bf G}_0\right)$$
$$-({\bf H}_2\cdot\nabla_{\bf x}){\bf v}+({\bf v}\cdot\nabla_{\bf y}){\bf H}_1$$
\begin{equation}
-\nabla_{\bf y}\times\left(\sum_{k=1}^3({\bf v}\times{\bf S}_k)H^k_1
+\sum_{k=1}^3\sum_{m=1}^3\lb{\bf v}\times{\bf\Gamma}_{mk}\rb
{\partial H^k_0\over\partial y_m}\right)+\lambda_2{\bf G}_0.
\label{eigeneq2osc}\end{equation}
Here, only the terms involving ${\bf H}_2$ and $\partial H^k_1/\partial y_m$
are not yet determined. However, as one can immediately see,
$\bf x$-dependent prefactors in front of these terms
are identical to those, which are in \rf{eigeneq1osc} in front of
${\bf H}_1$ and $\partial H^k_0/\partial y_m$, respectively.
This implies a representation
$$
{\bf G}_2=\sum_{k=1}^3{\bf S}_k({\bf x})H^k_2({\bf y})+
\sum_{k=1}^3\sum_{m=1}^3{\bf\Gamma}_{mk}({\bf x})
{\partial H^k_1\over\partial y_m}({\bf y})+{\bf Q}_2({\bf x},{\bf y}),
$$
where the vector field ${\bf Q}_2$ can be uniquely determined from
$${\cal L}{\bf Q}_2=-\eta\left(2\sum_{k=1}^3\sum_{m=1}^3\sum_{l=1}^3
{\partial{\bf\Gamma}_{mk}\over\partial x_l}
{\partial^2H^k_0\over\partial y_m\partial y_l}
+\nabla^2_{\bf y}{\bf G}_0\right)$$
$$
-\nabla_{\bf y}\times\sum_{k=1}^3\sum_{m=1}^3\lb{\bf v}\times{\bf\Gamma}_{mk}\rb
{\partial H^k_0\over\partial y_m}+\lambda_2{\bf G}_0.
$$

$iv$. As already stated, $\lambda_2$ is one of the two eigenvalues of
\rf{3deigeneq}-\rf{3dsolenoidality}. In what follows we assume
that the other eigenvalue, $\lambda_2'$, is distinct from $\lambda_2$
and we denote by ${\bf h}'$ the associated eigenvector.

Equations arising at higher orders in $\epsilon$ are now solved recursively.
Assume that all the equations up to order $\epsilon^{N-1}$ have been
solved, so that the following information has been obtained:

\noindent
$\bullet$ vector fields ${\bf H}_n$ and ${\bf G}_n$ for all $n<N-2$;

\noindent
$\bullet$ representations of ${\bf G}_n$ of the form
\begin{equation}
{\bf G}_n=\sum_{k=1}^3{\bf S}_k({\bf x})H^k_n({\bf y})+
\sum_{k=1}^3\sum_{m=1}^3{\bf\Gamma}_{mk}({\bf x})
{\partial H^k_{n-1}\over\partial y_m}({\bf y})+{\bf Q}_n({\bf x},{\bf y})
\label{Gn}\end{equation}
for $n=N-1$ and $n=N-2$ with known vector fields
${\bf Q}_n$, $\la{\bf Q}_n\ra=0$;

\noindent
$\bullet$ quantities $\lambda_n$ for $n<N$.

Upon substitution of \rf{Gn} for $n=N-1$ the average
of the equation at order $\epsilon^N$ becomes
$$
(\overline{\cal M}-\lambda_2){\bf H}_{N-2}-\lambda_N{\bf H}_0
=-\nabla_{\bf y}\times\la{\bf v}\times{\bf Q}_{N-1}\ra
+\sum_{m=1}^{N-3}\lambda_{N-m}{\bf H}_m,
$$
where the right-hand side at this stage is a known vector
field. Projecting this equation out in the direction of ${\bf H}_0$
one can uniquely determine $\lambda_N$. In the complementary invariant
subspace the operator $\overline{\cal M}-\lambda_2$ is invertible, and
hence ${\bf H}_{N-2}$ can be determined up to an arbitrary multiple of
${\bf H}_0$. This additive term can be neglected: one can demand that
${\bf H}_{N-2}$ belong to the complementary subspace. This is a
normalization condition related to the fact that eigenvectors can be
multiplied by arbitrary (analytic) functions of $\epsilon$. As a
result, ${\bf G}_{N-2}$ is now determined by \rf{Gn} with $n=N-2$.

Upon use of \rf{Gn} for $n=N-1$ the fluctuating part of the equation
at order $\epsilon^N$ becomes
$${\cal L}{\bf G}_N=-\eta\left(2\sum_{k=1}^3\sum_{m=1}^3\left(
{\partial{\bf S}_k\over\partial x_m}{\partial H^k_{N-1}\over\partial y_m}
+\sum_{l=1}^3{\partial{\bf\Gamma}_{mk}\over\partial x_l}
{\partial^2 H^k_{N-2}\over\partial y_l\partial y_m}\right)\right.$$
$$\left.+2(\nabla_{\bf x}\cdot\nabla_{\bf y}){\bf Q}_{N-1}
+\nabla^2_{\bf y}{\bf G}_{N-2}\right)-({\bf H}_N\cdot\nabla_{\bf x}){\bf v}
+({\bf v}\cdot\nabla_{\bf y}){\bf H}_{N-1}$$
$$-\nabla_{\bf y}\times\left(\sum_{k=1}^3({\bf v}\times{\bf S}_k) H^k_{N-1}
+\sum_{k=1}^3\sum_{m=1}^3\lb{\bf v}\times{\bf\Gamma}_{mk}\rb
{\partial H^k_{N-2}\over\partial y_m}+\lb{\bf v}\times{\bf Q}_{N-1}\rb\right)$$
\begin{equation}
+\sum_{m=0}^{N-2}\lambda_{N-m}{\bf G}_m.
\label{eigeneqNosc}\end{equation}
Only the terms involving ${\bf H}_N$ and derivatives of $H^k_{N-1}$
are not yet determined. Like in \rf{eigeneq2osc}, $\bf x$-dependent
prefactors in front of these terms are identical to those in front of
${\bf H}_1$ and derivatives of ${\bf H}^k_0$ in \rf{eigeneq1osc}, respectively.
Hence, \rf{eigeneqNosc} yields a representation of the form \rf{Gn} for
${\bf G}_N$. An equation for ${\bf Q}_N$ can be obtained from \rf{eigeneqNosc}
by replacing ${\bf G}_N$ by ${\bf Q}_N$ and omitting all terms involving
${\bf H}_N$ and derivatives of ${\bf H}_{N-1}$.

Thus, a complete formal decomposition of magnetic modes and their eigenvalues
has been constructed. A step-by-step analysis of the solution reveals that
$$
{\bf H}_0={\bf h}e^{i\bf q\cdot y},\quad{\bf H}_n=\chi_n{\bf h}'e^{i\bf qy}
\quad\forall n>0
$$
(with suitable scalars $\chi_n$) and
\begin{equation}
{\bf G}_n={\bf g}_n({\bf x})e^{i\bf q\cdot y}\quad\forall n\ge0;
\label{Gnexpform}\end{equation}
thus the expanded eigenmode admits a representation \rf{Hform} in the
form of a plane wave in the slow variable. (For this reason it was
sufficient to demand, for constructions presented under heading $iv$,
that the two eigenvalues $\lambda_2$ and $\lambda_2'$ of
$\overline{\cal M}$ be distinct in the subspace of solenoidal vector
fields ${\bf c}e^{i\bf qy}$ with ${\bf c}=$const, instead of requiring
that $\overline{\cal M}-\lambda_2$ be invertible in the whole domain.)
The plane-wave representation does not come as a surprise, since
the domain of the magnetic induction operator $\cal M$ splits into
invariant subspaces, each comprised of vector fields of the form
\rf{Hform} and categorized by wavevectors $\bf q$.

If the homogenized operator $\overline{\cal M}$ turns out to be
elliptic, one can prove, following Vishik (1986, 1987), that any point
$\lambda_2$ of the spectrum of $\overline{\cal M}$ is associated
with an analytic branch \rf{lambdaseries} of eigenvalues of the
original magnetic induction operator $\cal M$. The proof relies on
the general perturbation theory for linear operators (Kato, 1980); it will
not be presented here.

\section{Strongly negative eddy diffusivities}
\setcounter{equation}{0}
\label{a:loweddydiff}

In this Appendix we discuss a mechanism for appearance of negative
eddy diffusivities with arbitrarily large magnitudes.

Let ${\bf f}_p(\eta)$ denote the basis of small-scale
$2\pi$-periodic magnetic modes with vanishing mean and
$\zeta_p(\eta)$ -- the associated eigenvalues:
$$
{\cal L}{\bf f}_p(\eta)=\zeta_p(\eta){\bf f}_p(\eta),
\quad|{\bf f}_p(\eta)|=1,\quad\nabla\cdot{\bf f}_p(\eta)=0
$$
(assuming in this section, for the sake of simplicity, that
$\cal L$ is diagonalizable). Suppose that when molecular diffusivity $\eta$
is decreased, small-scale magnetic fields with vanishing mean lose
stability in such a way that the dominant eigenvalue of the magnetic
induction operator $\cal L$ passes through zero. Let
$\eta_c$ be the critical diffusivity:
$$
\hbox{Re }\zeta_p(\eta)<0\quad\forall p,\quad\forall\eta<\eta_c;
\quad\zeta_1(\eta_c)=0.
$$
Generically, the loss of stability occurs either in the proper
subspace comprised of parity-invariant fields, or in that of
parity-antiinvariant fields, implying two possible variants of the mechanism.

Suppose small-scale parity-invariant magnetic fields are the first to lose
stability. Solutions ${\bf S}(\eta)$ to the first auxiliary problem \rf{Seq}
continuously depend on $\eta$ for $\eta<\eta_c$. The second auxiliary
problem \rf{Gammaeq} can be reexpressed as
\begin{equation}
\sum_p\zeta_p(\eta)\gamma_{mk,p}(\eta){\bf f}_p(\eta)=
\sum_p\kappa_{mk,p}(\eta){\bf f}_p(\eta),
\label{newGammaeq}\end{equation}
where
$$
-2\eta{\partial{\bf S}_k(\eta)\over\partial x_m}
+v^m({\bf S}_k(\eta)+{\bf e}_k)-{\bf v}S^m_k(\eta)=
\sum_p\kappa_{mk,p}(\eta){\bf f}_p(\eta)
$$
and
$$
{\bf\Gamma}_{mk}(\eta)=\sum_p\gamma_{mk,p}(\eta){\bf f}_p(\eta)
$$
are decompositions of the r.h.s. of \rf{Gammaeq} and of ${\bf\Gamma}_{mk}(\eta)$
in the basis of magnetic modes; summation over all parity-invariant
eigenvectors is assumed in the two sums. From \rf{newGammaeq},
$\gamma_{mk,p}(\eta)=\kappa_{mk,p}(\eta)/\zeta_p(\eta)$.
Since generically the coefficients $\kappa_{mk,1}(\eta_c)$ do not
vanish, we have in the limit $\eta\to\eta_c$
$$
{\bf\Gamma}_{mk}(\eta)\approx{\kappa_{mk,1}(\eta_c)\over
\zeta_1(\eta)}{\bf f}_1(\eta_c)\to\infty.
$$
Then the equation \rf{3deigeneq} for the eigenvalue $\lambda_2$
asymptotically reduces to
\begin{equation}
{{\bf A}({\bf q})\over\zeta_1(\eta)}\sum_{k=1}^3\sum_{m=1}^3\kappa_{mk,1}
(\eta_c)h^kq^m=-\lambda_2{\bf h},
\label{ass1eigeneq}\end{equation}
where ${\bf A}({\bf q})\equiv{\bf q}\times\la{\bf v}\times{\bf f}_1(\eta_c)\ra$.
Equations \rf{ass1eigeneq} and \rf{3dsolenoidality} yield two eigenvalues.
It is easily checked that one of them, denoted $\lambda'_2$, vanishes and
thus is of no interest for us; the associated eigenvector ${\bf h}'$
is orthogonal to the wavevector $\bf q$ and satisfies
$$
\sum_{k=1}^3\sum_{m=1}^3\kappa_{mk,1}(\eta_c)h'^kq^m=0.
$$
The second eigenvalue is
$$
\lambda_2=-{a({\bf q})\over\zeta_1(\eta)},\qquad
a({\bf q})\equiv\sum_{k=1}^3\sum_{m=1}^3\kappa_{mk,1}(\eta_c)A^k({\bf q})q^m,
$$
the associated eigenvector being ${\bf h}={\bf A}({\bf q})$ (we
consider a generic case, where $\bf A$ does not vanish). Therefore,
if $\widehat{a}\equiv\max_{|{\bf q}|=1}a({\bf q})>0$, then, in the limit
$\eta\to\eta_c$, the minimum magnetic eddy diffusivity
becomes arbitrarily large negative:
$$
\eta_{\rm eddy}\approx{\widehat{a}\over\zeta_1(\eta)}\to-\infty.
$$

Alternatively, suppose now that the loss of stability occurs in the subspace
of small-scale parity-antiinvariant magnetic fields. Then the mechanism is
similar. In this case already the solutions ${\bf S}(\eta)$ to the
first auxiliary problem \rf{Seq} tend to infinity. Indeed,
\rf{Seq} can be represented as
$$
\sum_p\zeta_p(\eta)s_{k,p}(\eta){\bf f}_p(\eta)=
\sum_p\rho_{k,p}(\eta){\bf f}_p(\eta),
$$
where
$$
-{\partial{\bf v}\over\partial x_k}=\sum_p\rho_{k,p}(\eta){\bf f}_p(\eta)
$$
and
$$
{\bf S}_k(\eta)=\sum_ps_{k,p}(\eta){\bf f}_p(\eta)
$$
are decompositions of the r.h.s. of the first auxiliary problem \rf{Seq}
and of ${\bf S}_k(\eta)$ in the basis of magnetic modes; summation over
parity-antiinvariant eigenvectors is assumed in both sums. Thus,
$s_{k,p}(\eta)=\rho_{k,p}(\eta)/\zeta_p(\eta)$. Since generically
$\rho_{k,1}(\eta_c)\ne0$, we have in the limit $\eta\to\eta_c$
$$
{\bf S}_k(\eta)\approx{\rho_{k,1}(\eta_c)\over\zeta_1(\eta)}
{\bf f}_1(\eta_c)\to\infty.
$$
Let $\bgm_m$ denote the solution to
$$
{\cal L}\bgm_m=-2\eta_c{\partial{\bf f}_1(\eta_c)\over\partial x_m}
+v^m{\bf f}_1(\eta_c)-{\bf v}f^m_1(\eta_c).
$$
Then
$$
{\bf\Gamma}_{mk}(\eta)\approx{\rho_{k,1}(\eta_c)\over\zeta_1(\eta)}\bgm_m,
$$
and \rf{3deigeneq} asymptotically reduces to
\begin{equation}
{{\bf B}({\bf q})\over\zeta_1(\eta)}\sum_{k=1}^3\rho_{k,1}(\eta_c)h^k
=-\lambda_2{\bf h},
\label{ass2eigeneq}\end{equation}
where
$$
{\bf B}({\bf q})\equiv\sum_{m=1}^3q^m{\bf q}\times\la{\bf v}\times\bgm_m\ra.
$$
Equations \rf{ass2eigeneq} and \rf{3dsolenoidality} yield two eigenvalues.
Again, one can easily check that one of them, $\lambda'_2$, vanishes;
the associated eigenvector ${\bf h}'$ is orthogonal to $\bf q$ and satisfies
$$
\sum_{k=1}^3\rho_{k,1}(\eta_h)h'^k=0.
$$
The second is
$$
\lambda_2=-{b({\bf q})\over\zeta_1(\eta)},\qquad
b({\bf q})\equiv\sum_{k=1}^3\rho_{k,1}(\eta_c)B^k({\bf q}),
$$
with the associated vector ${\bf h}={\bf B}({\bf q})$ (in the generic case
${\bf B}\ne0$). Therefore, if $\widehat{b}\equiv\max_{|{\bf q}|=1}b({\bf q})>0$,
when $\eta\to\eta_c$ the minimum magnetic eddy diffusivity becomes
again arbitrarily large negative:
$$
\eta_{\rm eddy}\approx{\widehat{b}\over\zeta_1(\eta)}\to-\infty.
$$

\section{Chebyshev iterative methods
for numerical solution of systems of equations}
\setcounter{equation}{0}
\label{a:chebyshev}

We discuss here three Chebyshev iterative methods suitable for large
systems of linear or non-linear equations. They employ extremal
properties of Chebyshev polynomials and their roots. The first method
(\cu), discussed in \xsb{subs:old}, is applicable to problems with
negative-definite self-adjoint Jacobians. An extension (\cd),
discussed in \xsb{subs:mod}, is an upgrade of \cu, which eliminates
these restrictions. The second method (\ct), discussed in
\xsb{subs:new}, is designed for discretizations of elliptic
differential equations, such as the auxiliary problems \rf{Seq} and
\rf{Gammaeq}. \cu\ and \cd\ were introduced by Podvigina and
Zheligovsky (1997) and \ct\ is presented in detail in Zheligovsky
(2001). For general background on Chebyshev iterative methods, see
Axelsson (1996).

\subsection{The basic method: restricted case (\cu)}
\label{subs:old}

We consider numerical solution of a system of equations
\begin{equation}
{\bf F}({\bf z})=0,
\label{steadyeq}\end{equation}
where ${\bf F}:{\bf C}^N\to{\bf C}^N$ is smooth
and the number of equations, $N$, is large.

Let $\bf z=Z$ be a solution to \rf{steadyeq}. We first
consider the case, where the Jacobian matrix ${\bf F'}({\bf Z})$
is diagonalizable and its spectrum is (real and)
negative. Then the solution can be computed by iterations
\begin{equation}
{\bf z}_1={\bf z}_0+h{\bf F}({\bf z}_0),
\label{firststit}\end{equation}
\begin{equation}
{\bf z}_{k+1}=\gamma_k\left({\bf z}_k+h
{\bf F}({\bf z}_k)\right)+(1-\gamma_k){\bf z}_{k-1},\quad k>0,
\label{kthit}\end{equation}
provided ${\bf z}_0$ is chosen in the basin of attraction of $\bf Z$. Here
\begin{equation}
h\equiv{2\over\Lambda+\lambda},\qquad
\gamma_k\equiv2\mu{T_k(\mu)\over T_{k+1}(\mu)},\qquad
\mu\equiv{\Lambda+\lambda\over\Lambda-\lambda},
\label{step}\end{equation}
where $0\le\lambda<\Lambda$ for any $\Lambda\ge|{\bf F'}({\bf Z})|$
(the optimal value $\Lambda=|{\bf F'}({\bf Z})|$ is usually unknown),
and where the $T_k$'s are Chebyshev polynomials of the first kind.
The $\gamma_k$'s can be evaluated recursively:
\begin{equation}
\gamma_0=2;\quad\gamma_k={4\mu^2\over4\mu^2-\gamma_{k-1}},\quad k>0.
\label{gammakit}\end{equation}

Consider how the discrepancy vectors $\bzt_k\equiv{\bf z}_k-{\bf Z}$ evolve
in the course of iterations \rf{firststit}-\rf{gammakit}. Introduce
the eigenvalues $\eta_i$ of ${\bf F'}({\bf Z})$
(here,\break$0>\eta_i\ge-\Lambda$) and the associated eigenvectors $\{{\bf e}_i\}$,
and decompose the initial discrepancy in this basis:
$$
\bzt_0=\sum_{i=1}^N c^i{\bf e}_i.
$$
It is then easily shown that
$$
\bzt_k=\sum_{i=1}^N{T_k(\mu(1+h\eta_i))\over T_k(\mu)}
\ c^i{\bf e}_i+o(|\bzt_0|),\quad k\ge0
$$
(the $o(|\bzt_0|)$ correction does not appear if the mapping $\bf F$ is linear).

At this point, the standard strategy is to use the least deviation
principle for Chebyshev polynomials: the iterative process
\rf{firststit}-\rf{gammakit} with $\lambda>0$ ensures maximum
uniform decay of the discrepancy vector component in the proper
subspace of ${\bf F'}({\bf Z})$ associated with the part of the
spectrum in the interval $-\lambda\ge\eta_i\ge-\Lambda$. The
alternative procedure used by Podvigina and Zheligovsky (1997) relies,
in addition, on the extremal property of the largest root of a
Chebyshev polynomial: for $\lambda=0$, the iterative process gives
optimal reduction of the discrepancy vector component in the proper
subspace of ${\bf F'}({\bf Z})$ associated with eigenvalues $\eta_i$
closest to zero. Therefore, a suitable algorithm, called here \cu,
consists of a succession of iterative sequences
\rf{firststit}-\rf{gammakit} with $\lambda>0$ and $\lambda=0$.

As usually, lowering of the condition number of the system allows to improve
efficiency of the algorithm. If, in addition, ${\bf F'}({\bf Z})$ is
self-adjoint, such lowering can be achieved by replacing \rf{steadyeq}
with an equivalent system
$$
\xht{F}({\bf z})=0,
$$
where
\begin{equation}
\xht{F}({\bf z})\equiv{\bf P}({\bf z})\cdot{\bf F}({\bf z}),
\label{simplenewmap}\end{equation}
${\bf P}({\bf z})$ being a self-adjoint and positive-definite linear
preconditioning operator. It may be checked that for such a
${\bf P}({\bf z})$ the mapping $\xht{F}$ has the required properties stated
at the beginning of this subsection.

\subsection{An extension of \cu: general case (\cd)}
\label{subs:mod}

In the case of an arbitrary smooth mapping $\bf F$, to which
we turn now, a modified system
$$
\tld{F}({\bf z})=0
$$
can be considered, where
\begin{equation}
\tld{F}({\bf z})\equiv-{\bf P}_1({\bf z})\cdot({\bf F'}({\bf z}))^*
\cdot{\bf P}_2({\bf z})\cdot{\bf F}({\bf z}),
\label{newmap}\end{equation}
${\bf P}_i\ (i=1,2)$ being arbitrary positive definite self-adjoint operators
and\break${\bf M}^*=\overline{\bf M}^t$ being the operator adjoint to $\bf M$.
The method discussed above can be applied for the solution of the modified
system, since $\tld{F}'({\bf Z})$ has all the required properties.

Termination of individual sequences and selection of values of $\lambda$
in successive sequences is controlled by the following rules:

\noindent 1. In the first iterative sequence $\lambda=\lambda_1$.

\noindent 2. In any of the following iterative sequences $\lambda=0$,
unless $\lambda$ was zero in the previous sequence and that sequence
was short (i.e. it consisted of less than
$\left[\pi\sqrt{\Lambda\over8\lambda_1}+1\right]$ iterations, where
$[\cdot]$ denotes the integer part of a number), in which case in the
next sequence $\lambda=\lambda_1$.

\noindent 3. An iterative sequence with $\lambda=\lambda_1$ can be
terminated only after a prescribed number of iterations have been performed.

\noindent 4. In an iterative sequence with $\lambda=0$ the
errors $|{\bf F}({\bf z}_k)|$ can
initially grow monotonically, if the respective errors $|\tld{F}({\bf
z}_k)|$ decrease monotonically. (An increase of $|\tld{F}({\bf z}_k)|$
at the first iteration indicates that $\Lambda$ is underestimated.)

\noindent 5. After the initial phase allowed by rules 3 or 4 is completed, an
iterative sequence is terminated if both discrepancy reduction rate values,
${1\over k}\ln{|{\bf F}({\bf z}_0)|\over|{\bf F}({\bf z}_k)|}$ and
${1\over k}\ln{|\tld{F}({\bf z}_0)|\over|\tld{F}({\bf z}_k)|}$, are
smaller than at the previous iteration.

\noindent 6. Either the last or the next to last iterate is accepted as the
output of a sequence -- specifically, the one, for which the error
$|{\bf F}({\bf z}_k)|$ is smaller.

With these rules, the behavior of the algorithm proved insensitive to
the particular choice of $\lambda_1$ in the interval
$0.1\Lambda\le\lambda_1\le0.5\Lambda$.

Only trivial changes to this set of rules are required to make it applicable
to \cu.

\cd\ proved robust and efficient, e.g.\ in computation
of steady states of the Navier--Stokes equation (Podvigina, 1999a,b).
However, the use of \rf{newmap} may be potentially disadvantageous, because:

\noindent
$\bullet$ The presence of the adjoint to ${\bf F'}({\bf z})$ leads to
an increase in the condition number, which may not be completely neutralized
by the preconditioning operators ${\bf P}_1$ and ${\bf P}_2$.

\noindent
$\bullet$ Computation of $\tld{F}$ involves more operations
(and hence larger codes) than that of the original mapping $\bf F$.

\noindent
$\bullet$ Spurious roots may appear.

\subsection{An adaptation of \cu\ to elliptic PDE's (\ct\,)}
\label{subs:new}

We present here an alternative to \cd, better suited to problems
arising from discretization of elliptic partial differential
equations, when they involve not just leading-order elliptic operators
(such as the Laplacian), but also lower-order non-self-adjoint
perturbations. We do not attempt to present this method in full
generality (see Zheligovsky, 2001). In what follows the resolution of
the discretization, assumed to be large, is the control parameter,
which we denote by $R$.

Construction of the modified mapping \rf{newmap} can be avoided under
the conditions, that the basic mapping ${\bf F}$ can be split:
\begin{equation}
{\bf F}({\bf z})={\bf D}({\bf z})+{\bf A}({\bf z})
\label{decomp}\end{equation}
and if a linear invertible preconditioning operator $\bf P$ can be selected,
such that:

1. All eigenvalues $\sigma_i$ of the linear operator
${\cal D}\equiv{\bf P\cdot D'}({\bf Z})$
are negative. (If $\bf P$ is a self-adjoint positive-definite
operator and ${\bf D'}({\bf Z})$ is a self-adjoint negative-definite
operator, this condition is automatically verified.)

2. The ratio $\delta\equiv|{\cal A}|/\Lambda$ is small for large $R$. Here
${\cal A}\equiv{\bf P\cdot A'}({\bf Z})$, and $\Lambda$ is a ``tight''
upper bound for the spectral radius of $\cal D$, i.e.\ a bound, which
stays of the order of the spectral radius for large resolutions.

3. The operators ${\cal D}$ and ${\cal A}$ nearly commute:
$|{\cal DA}-{\cal AD}|=o(|{\cal DA}|)$.
(Note, that this condition is not that restrictive: it holds automatically in
the absence of preconditioning, if ${\cal D}$ is for example the Laplacian.)

A decomposition \rf{decomp} arises naturally if the problem \rf{steadyeq}
is a discretization of a dissipative partial differential equation: then
the mapping $\bf D$ can be identified with the higher-order elliptic part
of the equation, and $\bf A$ -- with the remaining terms. Auxiliary problems
\rf{Seq} and \rf{Gammaeq} arising from the two-scale expansion fall into
this class.

Under the conditions stated above a modification of the method
discussed in \xsb{subs:old}, named \ct, can be applied. Solution
proceeds by a succession of Chebyshev iterative sequences
\rf{kthit}-\rf{gammakit}, but the first iteration in all sequences is made
differently. Suppose a termination condition (for example, the one considered
in the previous subsection) for a Chebyshev sequence number $n-1$ is
verified for the first time in this sequence for the iterate ${\bf
z}^{n-1}_K$ (we then call $K$ the termination signal number). Then
the sequence is continued by $M$ further iterations, and we start the
next sequence, by selecting
$$
{\bf z}^n_0={\bf z}^{n-1}_K;\quad{\bf z}^n_1={\bf z}^{n-1}_K+h\left(
{\cal D}({\bf z}^{n-1}_K)+{\cal A}({\bf z}^{n-1}_{K+M})\right).
$$
Subsequent iterations are defined as in \rf{kthit}:
$$
{\bf z}^n_{k+1}=\gamma_k\left({\bf z}^n_k+h\xht{F}({\bf z}^n_k)
\right)+(1-\gamma_k){\bf z}^n_{k-1},\quad k>0.
$$
(The quantities $h,\ \mu$ and $\gamma_k$ were
defined by \rf{step}-\rf{gammakit}
and $\xht{F}$ -- by \rf{simplenewmap}.)

In this iterative process, errors behave in the following
way. Discrepancy vectors $\bzt^n_k\equiv{\bf z}^n_k-{\bf Z}$ can be
represented in terms of Chebyshev polynomials of the first ($T_k$) and
of the second ($U_k$) kind:
$$
\bzt^n_k={1\over
T_k(\mu)}(T_k({\cal Q}){\bf a}^n+U_k({\cal Q}){\bf b}^n)+o(|\bzt^n_0|),
$$
where ${\cal Q}\equiv\mu({\bf I}+h({\cal D}+{\cal A}))$, $\bf I$ is the
identity operator, and vectors ${\bf a}^n$ and ${\bf b}^n$, which we call the
basic error vectors, can be determined
from $\bzt^n_0$ and $\bzt^n_1$. (The terms $o(|\bzt^n_k|)$ do not
emerge if the mapping $\bf F$ is linear; in what follows they are neglected.)

Now, suppose that for the ($n-2$)-nd sequence the termination signal number
$K$ was the same as for the ($n-1$)-st iteration
(in computations, successive Chebyshev
sequences often do have equal lengths). Then it can be shown that

$\bullet$ $|{\bf b}^n|=O(\delta|{\bf a}^{n-1}|)$ and thus convergence of
${\bf b}^n$ is slaved to convergence of ${\bf a}^n$.

$\bullet$ Within the succession of iterative sequences, the basic
error vectors evolve as follows:
\begin{equation}
{\bf a}^n=\sum_{i=1}^N{a^{n-1}_i\over T_K(\mu)}(T_K(\xi_i){\bf e}_i
+h\mu f_i\,T'_K(\xi_i){\cal A}{\bf e}_i)+o(\delta|{\bf a}^{n-1}|),
\label{andecomp}\end{equation}
where
$$
f_i\equiv1+{1\over K}\left({T_K(\mu)\over T_{K+M}(\mu)}
{T_{K+M}(\xi_i)\over T_K(\xi_i)}-1\right),
$$
$\xi_i\equiv\mu(1+h\sigma_i)$, $\sigma_i$ being eigenvalues of $\cal D$,
and the coefficients $a^{n-1}_i$ can be found from the decomposition
of ${\bf a}^{n-1}$ in the basis of eigenvectors ${\bf e}_i$ of $\cal D$:
$$
{\bf a}^{n-1}=\sum_{i=1}^Na^{n-1}_i{\bf e}_i.
$$

Thus, reduction of the basic error vectors ${\bf a}^n$ is predominantly
controlled by $\cal D$ (note that $h|{\cal A}|=O(\delta)$). The
standard first Chebyshev iteration
\rf{firststit} is recovered when one sets $M=0$ and hence $f_i=1$.
It may be shown that our procedure leads to factors $0<f_i<1$, and
thus to enhanced convergence, for eigenvalues sufficiently small in
absolute value (provided $\lambda=0$ or $\lambda$ is small). For the
other eigenvalues good convergence is ensured as usually. (Actually,
because of the presence of the non-self-adjoint perturbation $\cal A$,
fast convergence is guaranteed only for eigenvalues large in absolute
value; for those of intermediate magnitude slowdown might occur, but has
not been observed in any of the cases we studied.)
Of course, these convergence arguments hold only in so far as the resolution
$R$ is high enough and thus $\delta$ is small enough
to justify keeping only the leading terms in \rf{andecomp}.

In the application of the generalized Chebyshev method to the
auxiliary problems \rf{Seq} and \rf{Gammaeq}, we take ${\bf D}=\eta\nabla^2$
(more precisely, $D$ is the Fourier--Galerkin
discretization of this operator). We select the preconditioning operator
${\bf P}=(-{\bf D})^{-q}$ for some $0<q<1$ by analogy with Podvigina
and Zheligovsky (1997) and Podvigina (1999a,b).
Then ${\cal D}=-(-{\bf D})^{1-q}$ is a
pseudodifferential operator of order $2-2q$, and $\cal A$ -- of
order $1-2q$. Hence, Condition 2 is inherited from the
non-discretized versions of the operators for $q<1$. Condition 3 is
satisfied for $q<3/4$. Indeed, in the commutator ${\cal DA}-{\cal AD}$ the
leading symbols cancel out and therefore its order is by 1 less than
the order of each of the operators $\cal DA$ and $\cal AD$; thus
$|{\cal DA}-{\cal AD}|=O(R^{\max(0,2-4q)})$,
$|{\cal DA}|=O(R^{\max(0,3-4q)})$, where $R$ is the maximum wavenumber of the
Fourier harmonics, retained in the abridged series, this completing
the argument. Since Condition 3 is also trivially satisfied for
$q=1$, it is natural to consider the larger interval $0\le q\le1$,
and we find the convergence to be the best for $q=3/4$. When applied to
\rf{Seq} and \rf{Gammaeq}, \ct\ proved superior to \cd.

\pagebreak

\end{document}